\documentclass[sigconf, nonacm]{acmart}

\newcommand\vldbdoi{XX.XX/XXX.XX}

\newcommand\vldbvolume{15}
\newcommand\vldbissue{13}

\newcommand\vldbavailabilityurl{https://github.com/utsaslab/dinomo}
\newcommand\vldbpagestyle{empty}

\def\showcomments{0}
\def\removespace{0}
\def\camerareadyversion{1}

\usepackage{tikz}
\usepackage{amsmath}

\usepackage{pifont}

\usepackage{booktabs}

\usepackage{filecontents}

\usepackage{paralist}
\usepackage{xspace}
\usepackage{tabularx}
\usepackage{subfigure}
\usepackage{multirow}
\usepackage{enumitem}
\usepackage[utf8]{inputenc}

\usepackage[nameinlink]{cleveref}

\begin{document}
\newcommand{\mycaption}[2]{\caption{\textbf{#1}. {#2}}}
\newcommand{\sref}[1]{\S\ref{#1}}
\newcommand{\vheading}[1]{\vspace{0.05in}\noindent\textbf{#1}}
\newcommand{\viheading}[1]{\vspace{0.05in}\noindent\emph{#1}}
\newcommand{\fsync}{\texttt{fsync()}\xspace}
\newcommand{\etal}{\textit{et al.}\xspace}
\newcommand{\ie}{\textit{i.e.,}\xspace}
\newcommand{\eg}{\textit{e.g.,}\xspace}
\newcommand{\etc}{\textit{etc.}\xspace}
\newcommand*{\affaddr}[1]{#1} 
\newcommand*{\affmark}[1][*]{\textsuperscript{#1}}
\newcommand{\myx}{$\times$\xspace}
\newcommand{\vtt}[1]{\texttt{#1}\xspace}
\newcommand{\blink}{$B^{link}$\xspace}
\newcommand{\fastfair}{FAST \& FAIR\xspace}
\newcommand{\lget}{\texttt{get()}\xspace}
\newcommand{\cas}{\texttt{CAS}\xspace}
\newcommand{\lput}{\texttt{put()}\xspace}
\newcommand{\lrq}{\texttt{range\_query}\xspace}
\newcommand{\lseek}{\texttt{seek()}\xspace}
\newcommand{\xtimes}{$\times$\xspace}
\newcommand{\store}{\texttt{store}\xspace}
\newcommand{\load}{\texttt{load}\xspace}
\newcommand{\rand}{\texttt{rand}\xspace}
\newcommand{\stride}[1]{\texttt{stride-{#1}}\xspace}
\newcommand{\llr}[1]{\texttt{linked-list-{#1}}\xspace}
\newcommand{\clflush}{\texttt{clflush}\xspace}
\newcommand{\clflushopt}{\texttt{clflushopt}\xspace}
\newcommand{\clwb}{\texttt{clwb}\xspace}
\newcommand{\filewb}{\texttt{file-writeback}\xspace}
\newcommand{\writeback}{\texttt{writeback}\xspace}
\newcommand{\sysnormal}{Recipe\xspace}
\newcommand{\stateart}{{state-of-the-art}\xspace}

\newcommand{\rmv}[1]{}
\newcommand{\revisit}[1]{}
\newcommand{\red}[1]{\textcolor{red}{#1}}

\newcommand{\proto}{{\textsc{Dinomo}}\xspace}
\newcommand{\sysname}{{\textsc{Dinomo}}\xspace}
\newcommand{\sysnametitle}{{\textsc{Dinomo}}\xspace}

\if\showcomments 1
\newcommand{\kim}[1]{{\textcolor{red}{Kim: #1}}}
\newcommand{\mka}[1]{{\textcolor{blue}{Marcos: #1}}}
\newcommand{\ssi}[1]{{\textcolor{green}{#1}}}
\newcommand{\sou}[1]{{\textcolor{violet}{Soujanya: #1}}}
\newcommand{\skl}[1]{{\textcolor{orange}{Sekwon: #1}}}
\newcommand{\sharad}[1]{{\textcolor{purple}{Sharad: #1}}}
\newcommand{\vijay}[1]{{\textcolor{blue}{Vijay: #1}}}
\else
\newcommand{\kim}[1]{}
\newcommand{\mka}[1]{}
\newcommand{\ssi}[1]{}
\newcommand{\sou}[1]{}
\newcommand{\skl}[1]{}
\newcommand{\sharad}[1]{}
\newcommand{\vijay}[1]{}
\fi

\newcommand{\dip}{{\textsc{OP}}\xspace}
\newcommand{\dpart}{ownership partitioning\xspace}
\newcommand{\dpartc}{Ownership partitioning\xspace}
\newcommand{\dpartcc}{Ownership Partitioning\xspace}

\newcommand{\dpm}{{DPM}\xspace}
\newcommand{\dpmfull}{{disaggregated PM}\xspace}
\newcommand{\sdpm}{{Disaggregated PM}\xspace}

\newcommand{\cptitle}{DAC\xspace}
\newcommand{\cp}{{\textsc{DAC}}\xspace}
\newcommand{\cpfulllower}{disaggregated adaptive caching\xspace}
\newcommand{\cpfull}{{\small\textsc{Disaggregated Adaptive Caching}}\xspace}
\newcommand{\cpplain}{Disaggregated Adaptive Caching\xspace}
\newcommand{\cpingfull}{{\small\textsc{Disaggregated Adaptive Caching}}\xspace}
\newcommand{\cpingplain}{Disaggregated Adaptive Caching\xspace}

\newcommand{\cn}{{KN}\xspace}
\newcommand{\cns}{{KNs}\xspace}
\newcommand{\cnfull}{{KVS node}\xspace}
\newcommand{\cnsfull}{{KVS nodes}\xspace}
\newcommand{\cnsfullc}{{KVS nodes}\xspace}

\newcommand{\rn}{{RN}\xspace}
\newcommand{\rns}{{RNs}\xspace}
\newcommand{\rnfull}{{routing node}\xspace}
\newcommand{\rnsfull}{{routing nodes}\xspace}

\newcommand{\mmnode}{{M-node}\xspace}
\newcommand{\mmnodefull}{{monitoring/management node}\xspace}

\title{DINOMO: An Elastic, Scalable, High-Performance\\Key-Value Store
for Disaggregated Persistent Memory\\(Extended Version)}


\author{Sekwon Lee}
\affiliation{%
  \institution{University of Texas at Austin}
}
\if\camerareadyversion 1
\email{sklee@cs.utexas.edu}
\fi

\author{Soujanya Ponnapalli}
\affiliation{%
  \institution{University of Texas at Austin}
}
\if\camerareadyversion 1
\email{soujanyap95@gmail.com}
\fi

\author{Sharad Singhal}
\affiliation{%
  \institution{Hewlett Packard Labs}
}
\if\camerareadyversion 1
\email{sharad.singhal@hpe.com}
\fi

\author{Marcos K. Aguilera}
\affiliation{%
  \institution{VMware Research Group}
}
\if\camerareadyversion 1
\email{maguilera@vmware.com}
\fi

\author{Kimberly Keeton}
\affiliation{%
  \institution{Google}
}
\if\camerareadyversion 1
\email{kimberly.keeton@gmail.com}
\fi

\author{Vijay Chidambaram}
\affiliation{%
  \institution{University of Texas at Austin}
}
\affiliation{%
  \institution{VMware Research Group}
}
\if\camerareadyversion 1
\email{vijayc@utexas.edu}
\fi

\begin{abstract}

We present \sysname, a novel key-value store for 
disaggregated persistent memory (\dpm).
\sysname is the first key-value store for \dpm that simultaneously achieves high common-case performance, scalability, and lightweight online reconfiguration. 
We observe that previously proposed key-value stores for \dpm had architectural limitations that prevent them from achieving all three goals simultaneously. 
\sysname uses a novel combination of techniques such as \dpart, \cpfulllower, selective replication, and lock-free and log-free indexing to achieve these goals. 
Compared to a state-of-the-art \dpm key-value store,
\sysname achieves at least 3.8\myx better throughput on various workloads at scale and higher scalability, while providing fast reconfiguration. 
\end{abstract}

\maketitle

\setcounter{page}{1}

\pagestyle{\vldbpagestyle}
\if\camerareadyversion 1
\begingroup
\renewcommand\thefootnote{}\footnote{\noindent
This work is licensed under the Creative Commons BY-NC-ND 4.0 International License. Visit \url{https://creativecommons.org/licenses/by-nc-nd/4.0/} to view a copy of this license. For any use beyond those covered by this license, obtain permission by emailing \href{mailto:info@vldb.org}{info@vldb.org}. Copyright is held by the owner/author(s). Publication rights licensed to the VLDB Endowment. \\
\raggedright Proceedings of the VLDB Endowment, Vol. \vldbvolume, No. \vldbissue\ %
ISSN 2150-8097. \\
\href{https://doi.org/\vldbdoi}{doi:\vldbdoi} \\
}\addtocounter{footnote}{-1}\endgroup
\fi

\if\camerareadyversion 1
\ifdefempty{\vldbavailabilityurl}{}{
\vspace{.3cm}
\begingroup\small\noindent\raggedright\textbf{PVLDB Artifact Availability:}\\
The source code, data, and/or other artifacts have been made available at \url{\vldbavailabilityurl}.
\endgroup
}
\fi

\section{Introduction}

Large cloud providers operate at a much larger scale than traditional enterprise data centers and aim to optimize their infrastructures for high utilization.
However, recent work indicates that resources in cloud data centers remain underutilized~\cite{guz2017nof,AlibabaTrace-BIGDATA2017,BorgTrace-EuroSys2020,yizhouOSDI18}. 
In the face of dynamic and bursty workloads, scheduling tasks such that resource utilization is high proves challenging~\cite{zhangvldb21}.
For example, cluster memory utilization can be as low as 60\%~\cite{chengBigData18, VermaCluster14, TirmaziEurosys20}.

One promising way to increase  resource utilization is to disaggregate resources~\cite{sebastian2020disappl,limISCA09,peterOSDI16,keeton2015machine}.
In a disaggregated cluster, resources such as CPU, memory, and storage are each collected into a separate central network-attached pool.
By sharing these resources across users and applications, utilization can be increased significantly.
Furthermore, each resource can be scaled up or down independently of the others: for example, memory can be added without the need to also add CPU or storage.
Such disaggregation has long been practiced for storage in the form of network-attached storage (NAS)~\cite{gibsonNAS} and Storage Area Networks (SAN)~\cite{khattar1999introduction}.
In this work, we take the idea one step further and consider a cluster where Persistent Memory is disaggregated.

Persistent Memory (PM) is a new memory technology that provides durability like traditional storage, with performance close to DRAM~\cite{pm-arxiv, yang20-pm,IntelPMM}. Since PM has much higher cost per GB than conventional storage~\cite{assise-osdi20}, it is critical to achieve high utilization in PM deployments. Similar to  traditional storage, the utilization of PM would increase from disaggregation. However, the DRAM-like latencies of PM make disaggregation challenging, since the network latency is 
an order of magnitude higher than PM latency. 

Disaggregated Persistent Memory (\dpm) is still under active research and development, and hence there are different kinds of \dpm to build upon.
In this work, we assume that \dpm is available as a centralized, reliable pool accessible via the network~\cite{keeton2017machine}.
We further assume that \dpm includes limited computational capability, as prior work shows such capability is critical for achieving good performance~\cite{ma2020asymnvm,tsai2020disaggre,MingFAST22}.

We are interested in using DPM to build persistent key-value stores (KVSs), which
  are critical pieces of software infrastructure.
The KVS consists of a number of \cnsfull (\cns) equipped with general-purpose processors, a relatively
 small amount of local DRAM, and high-performance network primitives like RDMA to access \dpm over the network~\cite{volosSOCC18}.
An ideal KVS for \dpm would have a number of properties: high common-case performance, scalability, and quick reconfiguration that allows handling failures, bursty workloads, and load imbalance efficiently.

Building a KVS that achieves all the goals
    simultaneously is challenging.
First, \cns incur expensive network round trips (RTs)
    for accessing data and metadata in \dpm.
Despite these overheads, the KVS must provide high
    performance.
Second, to benefit from independent scaling of \cns and PM,
    the KVS must be elastic and support lightweight
    reconfiguration of resources.
Finally, the KVS must provide scalable performance
    without bottlenecks due to load imbalance at \cns or
    from non-uniform workload patterns.

Prior \dpm KVSs~\cite{ma2020asymnvm,tsai2020disaggre} make design trade-offs that make these goals difficult to satisfy simultaneously. For example, AsymNVM~\cite{ma2020asymnvm} achieves high performance by adopting a shared-nothing architecture to enable high cache locality at \cns.
However, expensive data reorganization is needed when changing the number of \cns or rebalancing their load, thus limiting
elasticity and efficient load balancing. Similarly, Clover~\cite{tsai2020disaggre} supports straightforward load balancing and high elasticity using a shared-everything architecture where data is shared across \cns, and any \cn can handle any request. However, performance and scalability suffer as a result of poor cache locality and consistency overheads (including cache coherence, contention, and synchronization overheads due to sharing)
in the common case~\cite{porobicVLDB12}. 

In this work, we present \textbf{\sysname}, the first \dpm KVS
that simultaneously achieves high common-case performance, scalability, and lightweight online reconfiguration.
\sysname also provides linearizable reads and writes.
To achieve these goals, \sysname carefully adapts techniques from the storage research communities, including caching, \dpart, selective replication, and lock-free and log-free PM indexing. 

\vheading{Data organization on \dpm}. \sysname stores data and metadata on \dpm to enable concurrent and consistent access by all \cns.
Because \dpm is shared among all \cns, it functions as the source of ground truth in the system.
To enable consistent updates, data is written to \dpm in the form of log entries by the \cns. 
These log entries are asynchronously merged in order into the metadata index by the processors at the \dpm.
For its metadata index, DPM uses a concurrent PM index~\cite{lee2019recipe} which provides lock-free reads and log-free in-place-writes; the lock-free reads allow us to eliminate synchronization overheads between KNs and log-free in-place-writes allow DPM processors to concurrently update the metadata.

\vheading{\cpplain (\cp)}. 
Similar to other disaggregated systems, \sysname reduces network RTs by caching data and metadata in the local DRAM of each \cn. 
Data is cached by storing the key-value pair, and metadata is cached by storing a pointer to the data on \dpm (termed \textit{shortcuts}~\cite{tsai2020disaggre}).
To determine how best to divide the cache space between data and metadata,
\sysname uses \cp, a novel adaptive caching policy
  that actively maintains the right balance between caching
  values and shortcuts based on the workload patterns
  and available memory at \cns. 
\cp allows \sysname to make efficient use of the DRAM at \cns without making any assumptions about the workload.

\vheading{\dpartcc (\dip)}. While caching at the \cns can reduce network RTs, it can incur significant
  consistency overheads when \cns can share the same data.
To handle this concern, \sysname partitions the \emph{ownership} of data across \cns, while data and metadata are shared via \dpm. 
This provides three benefits.
First, it allows \cns to cache the data they own, thus providing high cache locality without consistency overheads.
Second, by sharing the data and metadata, \dip supports changing the number of \cns or rebalancing their load by repartitioning only the ownership of data among \cns,
  without expensive data reorganization at \dpm.
Finally, since each key is  only accessed by one \cn at any given point, combined with our principled 
reconfiguration protocol, \sysname achieves linearizable reads and writes.
Similar ideas have been proposed before in other contexts~\cite{khattar1999introduction, caulfieldISCA13, snowflake, atulOSDI16}, but we are the first to adapt it for \dpm.
With \dip, \sysname achieves high performance/scalability from
    locality-preserving \cn-side caching
    without consistency overheads 
    and high elasticity from lightweight reconfiguration.

\vheading{Selective Replication}.  \sysname{}'s \dpart, however, may experience performance or scalability bottlenecks at \cns due to load imbalance under highly skewed workloads (\ie
  the maximum throughput for requests on a single key is limited
  by the processing capacity of a single \cn).
To avoid this issue and provide scalable performance for highly skewed workloads, \sysname \emph{selectively replicates} the ownership of hot keys
  across multiple \cns.
\sysname has a separate \mmnodefull
  that identifies hot keys, initiates their ownership
  replication to other \cns, and thus balances the load
  from hot keys across available \cns.



\vheading{Alleviate network and CPU bottleneck}.
\sysname{}'s data path uses \emph{one-sided RDMA operations} with \emph{asynchronous post-processing}. All reads to \dpm by \cns use one-sided RDMA operations on a shortcut hit or a cache miss.
\sysname writes multiple log entries in a batch in the critical path using a one-sided RDMA operation, and delegates the merging of the writes into the metadata index to the \dpm processors asynchronously.
Asynchronous post-processing reduces write latency and amortizes \dpm processing utilization across multiple writes, reducing how much \dpm computing power is needed in the critical path.
These optimizations decrease the network messages per operation and alleviate the processing bottleneck from \dpm, increasing the efficiency of \sysname in addition to techniques like \cp and \dip.

\vheading{Limitations}.
Our work has a number of limitations.
First, while we address the challenge of scaling \cns, we do not tackle how to make \dpm reliable or scalable. 
Second, \sysname{} targets key-value store functionality for \dpm systems.
Many of its ideas may be equally applicable for a broader range of \dpm-based storage systems as well as disaggregated DRAM systems, but we have not explored this.
Finally, while our work provides mechanisms for scaling \cns, it does not tackle the policy question of when \cns should be scaled.
We consider these areas ripe for future work.

\vheading{Evaluation}. We implement \sysname in 10K lines of C++ code.
 We compare the end-to-end performance and scalability of
    \sysname with Clover~\cite{tsai2020disaggre}, a state-of-the-art \dpm KVS.
Our experiments show that \sysname achieves both better common-case performance and scalability than Clover. 
\sysname's throughput scales to 16 \cns,
    while Clover's throughput does not scale beyond 4 \cns.
With 16 \cns, \sysname outperforms Clover by at least
    3.8\myx on all workloads we evaluate.
We also show that 
    \sysname elastically scales-out \cns,
    balances the load across \cns, and
    handles \cn failures quickly.

  

In summary, this paper makes the following contributions:
\begin{itemize}
    \item We present \sysname, the first \dpm key-value store that simultaneously achieves high performance, scalability, and lightweight online reconfiguration (\sref{sec:dinomo})

    \item We present \cp, a novel adaptive caching policy that helps utilize the \cn-side memory effectively without any assumptions on workload patterns (\sref{sec:adaptive-caching})

    \item We adapt \dip for \dpm KVSs
    to achieve 
    high performance, scalability, and
      lightweight reconfiguration (\sref{sec:ownership-part})


    
    \item We experimentally show that \sysname can efficiently react to both \cn failures and load imbalance, and automatically scale the number of \cns by capturing load dynamics (\sref{sec:evaluation})

\end{itemize}

\section{Background and Motivation}

\begin{table}[t]
\centering
\resizebox{.47\textwidth}{!}{%
\begin{tabular}{@{}lccc@{}}
\toprule
\multicolumn{1}{c}{\textbf{KVS property}} & \textbf{\sysname} & \textbf{Clover} & \textbf{AsymNVM} \\ \midrule
Data                                 & shared          & shared          & partitioned      \\
Metadata                             & shared          & shared          & partitioned      \\
Ownership of data           & partitioned     & shared          & partitioned         \\
\hline
High performance         & \checkmark             & \myx              & \checkmark              \\
Scalability         & \checkmark             & \myx              & \checkmark              \\
Lightweight reconfiguration          & \checkmark             & \checkmark             & \myx              \\ \bottomrule
\end{tabular}
}
\caption{Design choices and properties of different \dpm KVS}
\label{tab:kvs-comparison}
\if\removespace 1
\vspace{-20pt}
\fi
\end{table}

We describe persistent memory (PM) and how it can be used
    in disaggregated settings.
We then discuss prior key-value stores (KVSs) for disaggregated
    PM (\dpm) and motivate the need for a new KVS.
    
\subsection{Persistent Memory and Disaggregation}

PM is a non-volatile memory
    technology with unique characteristics~\cite{pm-arxiv, yang20-pm}.
PM is connected directly to the memory bus -- it is byte addressable,
    and has performance close to DRAM. 
It has high capacity: Intel's Optane DC PM is available up to 512GiB per NVDIMM~\cite{IntelPMM}.
The per-GB cost of PM is higher than high-end solid state drives,
    but less than DRAM~\cite{assise-osdi20}. 
To improve cost efficiency and PM utilization, 
    prior work
    proposes \dpm~\cite{tsai2020disaggre, ma2020asymnvm, MingFAST22, LiuICCD21, keetonOpenFAM19, volosSOCC18}.
    We note that our work is agnostic to the choice of PM technology and specific PM 
    product (\eg PCM~\cite{WongPCM}, STT-MRAM~\cite{ApalkovSTTMRAM}, Memristor~\cite{YangMemristor}, 
    Optane DC PM~\cite{IntelPMM}, Memory-Semantic CXL SSD~\cite{flash22-keynote}).

\vheading{Disaggregated PM}.
In disaggregated settings, PM is available as a central,
    reliable pool of memory accessible over a network.
\emph{\cnsfullc} (\cns) are used to access the data in \dpm;
\cns have limited DRAM and use network primitives like RDMA
    to access the PM pool over a fast interconnect such as
    InfiniBand~\cite{sebastian2020disappl},
    PMoF~\cite{Golander-SYSTOR17, Paul-PMSummit18},
    or Gen-Z~\cite{genz}.
Disaggregation allows independent scaling of PM and \cns
    and introduces separate failure domains, where
    \cn failures do not cause \dpm failures.

\dpm can be classified as active or passive. 
Active \dpm has small processing units such as ARM-SOCs,
     ASICs, or FPGAs, with high-bandwidth network ports.
In active \dpm, the compute capacity at \dpm
     is used for local processing, 
     including network, application-level, and data store processing~\cite{ma2020asymnvm, kim2018hyperloop, sidler2020strom}.
Prior work
    has proposed data stores for active \dpm that
    leverage this limited computational power~\cite{tsai2020disaggre, ma2020asymnvm, MingFAST22, zhiyuanASPLOS22, LiuICCD21}. 
    In contrast, passive \dpm has no computational abilities
     at the \dpm pool.
\cns can only use one-sided RDMA operations to 
    access and modify the data in \dpm.
Data stores for passive \dpm~\cite{tsai2020disaggre} have poor performance and scalability
    due to the limited functionality of the one-sided network primitives~\cite{aguilera2019designing},
    showing that active \dpm is a more practical deployment.

\subsection{DPM Key-Value Stores}\label{sec:back-mot-dpm-kvs}

Previously proposed \dpm key-values stores differ based on how they handle data, metadata, and ownership; metadata is information used to locate and access data (like an index); ownership determines if a data item can be read or written.

\vheading{AsymNVM}.
AsymNVM~\cite{ma2020asymnvm} adopts a shared-nothing architecture.
Data in \dpm is partitioned, and each partition is exclusively accessed by a single \cn.
Every \cn uses its local memory to cache data from its partition
    (Table~\ref{tab:kvs-comparison}); caching helps reduce
    expensive network round trips to \dpm.
As \cns have exclusive ownership over data,
    their caches can preserve high locality and can be consistent
    without incurring additional consistency overheads.
Thus, shared-nothing architectures provide high performance and scalability
    in the common case by effectively using \cn caches to
    process requests.
However, reconfiguring the number of \cns or balancing load across
    \cns requires physical reorganization of data and
    metadata~\cite{guz2017nof, klimovic2016flash, Bindschaedler2020hail, ma2020asymnvm}. 
For example, adding a new \cn may require the metadata of a
    partition to be split, resulting in expensive data copies at \dpm.
Thus, AsymNVM offers performance at the expense
    of elasticity and fast reconfiguration.

\vheading{Clover}.
Clover~\cite{tsai2020disaggre} adopts a shared-everything architecture.
All \cns share the ownership of data in \dpm, and
    every \cn can access and modify all data and metadata (Table~\ref{tab:kvs-comparison}).
\cns can use local memory to cache data.
However, due to sharing, \cns have poor cache locality and
    need to keep their caches consistent, incurring
    significant consistency overheads that reduce the
    common-case performance and scalability~\cite{porobicVLDB12}.
Nevertheless, Clover can support lightweight reconfiguration
    without re-partitioning data or metadata and
    allow straightforward load balancing across \cns.
Overall, Clover offers elasticity and lightweight reconfiguration
    at the expense of high common-case performance and scalability.

In summary, prior \dpm key-value stores sacrifice one of high common-case performance, scalability,
    or lightweight reconfiguration for the other two (Table~\ref{tab:kvs-comparison}). 
Thus, there is a need for a new \dpm key-value store that achieves
    the three properties simultaneously.

\section{\sysname}
\label{sec:dinomo}
\begin{table}[t]
\centering
\resizebox{.47\textwidth}{!}{%
\begin{tabular}{@{}l|c@{}}
\toprule
\textbf{Goals}   & \textbf{\sysname techniques}                      \\ \midrule

High performance & \dpartc, \cp                                    \\ \midrule
\begin{tabular}[c]{@{}l@{}}Lightweight reconfiguration\\ and scalability\end{tabular}
                 & \dpartc                                         \\ \midrule

  Linearizable reads and writes
                 & Shared \dpm, \dpartc              \\ 
    
  
                 \bottomrule
\end{tabular}
}
\caption{\sysname goals and design techniques}
\label{tab:design-summary}
\if\removespace 1
\vspace{-20pt}
\fi
\end{table}
We now present \sysname, a key-value store (KVS) for \dpm.
We first describe its API, target workloads, goals, and the guarantees it provides.
Then, we explain how \sysname achieves its goals (Table~\ref{tab:design-summary}).

\vheading{API}.
\sysname allows applications to perform \texttt{insert(key, value)}, \texttt{update(key, value)}, \texttt{lookup(key)},
    or \texttt{delete(key)} on variable-sized key-value pairs. 
We refer to the \texttt{lookup} operations as \emph{reads},
    and the \texttt{insert}, \texttt{update}, and \texttt{delete}
    operations as \emph{writes}. 
    
\vheading{Target workloads}. \sysname targets applications with dynamic working
    sets and sizes, and non-uniform workloads with varying
    skew~\cite{novakCLOUD20, quCSUR18, anna-vldb19}.
Large variations in workloads require \dpm KVSs to allow the
    elastic deployment of resources (e.g., \cns) in response to those dynamics~\cite{zhangvldb21,caoSIGMOD21}.
    
\vheading{Goals}. \sysname aims to achieve the following goals:
\begin{itemize}[leftmargin=10pt]
    \item High common-case performance in the absence of failures or reconfiguration
    \item Scalability of performance when the number of \cns increases
    \item Lightweight online reconfiguration
        to effectively handle \cn failures, bursty workloads, and
        load imbalance on available \cns
    \item Linearizable reads and writes 
\end{itemize}

\vheading{Guarantees}. \sysname guarantees that once committed, data will not be lost or corrupted regardless of \cns failures. 
It also ensures data remains available if at least one \cn and the \dpm are available. 

\subsection{Architecture}
\label{sec-arch}

\begin{figure}
  \centering
\includegraphics[width=.47\textwidth]{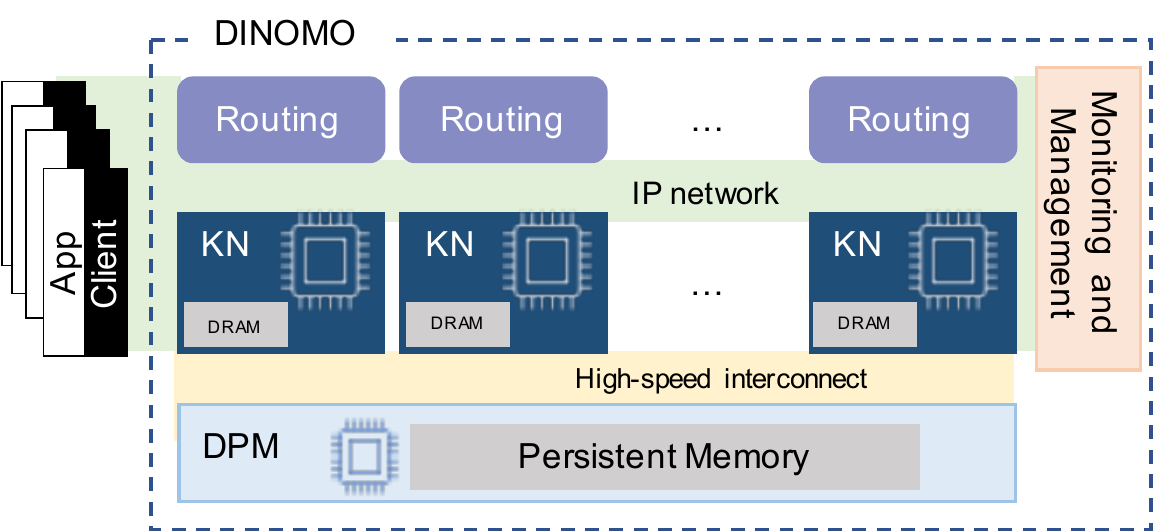}
\if\removespace 1
\vspace{-20pt}
\fi
\caption{Overview of the \sysname cluster}{}
\label{fig-high-level-arch}
\if\removespace 1
\vspace{-10pt}
\fi
\end{figure}

Figure~\ref{fig-high-level-arch} shows the high-level architecture of \sysname.
\sysname consists of clients, \rnsfull (\rns), \cnsfull (\cns), \dpm, and \mmnodefull (\mmnode).
We describe those components and how a request flows between them. 

Applications and users interact with \sysname through clients. 
\rns are the client-facing tier that provides cluster membership and isolate clients from the internal variance of the KVS cluster.
A client first contacts a \rn to obtain cluster membership and caches the mapping of key ranges to various \cns.
The client contacts the appropriate \cn that will then perform the read or write operation on its behalf. 
Each \cn is equipped with general-purpose processors and a small amount of DRAM relative to the \dpm capacity.
The \cn uses one-sided or two-sided RDMA primitives to access \dpm over the interconnect~\cite{sebastian2020disappl}; note that the one-sided RDMA primitive can read or write data without involving the \dpm processors. 
The \dpm has the large shared PM pool and limited computational power relative to \cns~\cite{ma2020asymnvm,tsai2020disaggre,MingFAST22}.
This asymmetry is deliberate: \cns are intended to run complex operations
    in a critical path, whereas the \dpm is intended to execute lightweight
    tasks outside the critical path, while keeping the cost of
    provisioning \dpm low.
The \cn caches the data it fetches from the \dpm in its local DRAM, and responds to the client request.
The \mmnode observes \cns statuses and workload characteristics to detect \cn failures, load imbalance, or workload skew, and triggers a suitable reconfiguration.

Note that we separately deploy the different functional components of \sysname to enable us to independently scale them up and down as required.
It is also possible to co-locate some components at the expense of reducing the efficiency of policy decisions when scaling resources.

\vheading{Assumptions}.
We assume all components in the \sysname cluster are inter-connected through a
    reliable local network (either over TCP/IP or RDMA RC).
The interconnect bandwidth between \cns and \dpm is lower than the memory bandwidth
    of the PM itself, usually making network the bottleneck~\cite{anujSOCC20, assise-osdi20}.
\cn failures are fail-stop and independent of \dpm failures; when an \cn fails,
    its DRAM contents are lost.
The \dpm has internal mechanisms or hardware support
    to ensure high availability~\cite{ma2020asymnvm, tsai2020disaggre,keetonOpenFAM19, lee2022hydra, zhouOSDI22}
    and hardware-level memory protection~\cite{nightingale2011cycles, sridharan2015memory,volosCAL21,zhangMICRO18}.
The \mmnode is always alive; this can be ensured via consensus and 
    replication~\cite{lamport2001paxos,lamport2019paxos,diegoATC14}.
As the \mmnode deals with infrequent lightweight tasks, using consensus does not introduce performance bottlenecks.

\subsection{Data organization on DPM}
\label{sec:dpm-sharing}
\begin{figure}
  \centering
\includegraphics[width=.46\textwidth]{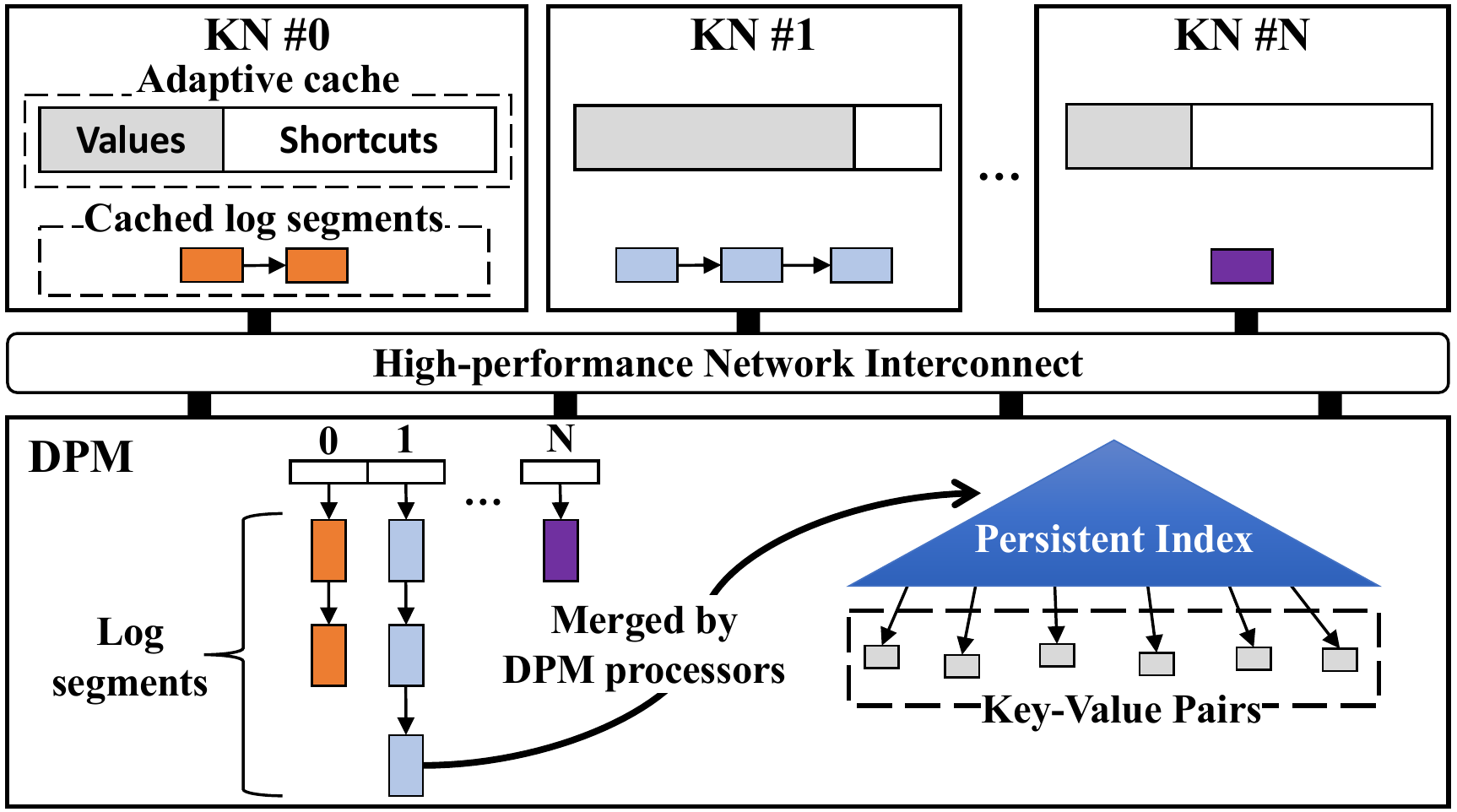}
\if\removespace 1
\vspace{-10pt}
\fi
\caption{\sysname Data Plane}{}
\label{fig-dinomo-data-plane}
\if\removespace 1
\vspace{-10pt}
\fi
\end{figure}

Figure~\ref{fig-dinomo-data-plane} shows the data-plane components in \sysname.
\sysname stores data (key-value pairs) and
    metadata (indexing structures) in \dpm,
    providing durability and the source of ground truth. 

\vheading{Storing data in logs}. 
In response to a write request, a \cn writes data to an exclusive log on \dpm.
This write is performed with a single one-sided write operation in the critical path.
The log is broken into a series of segments. 
Since each \cn handles requests on exclusive logical data partitions (\sref{sec:ownership-part}), two \cns will never log a write for the same key. 
The \dpm processors asynchronously
merge the write operations in a log segment in order into the metadata index.
Logs of different \cns may be merged into the index simultaneously. 
    
\vheading{Metadata index}. 
The metadata index in \dpm must satisfy
    the following requirements.
First, \cns should not hold locks while performing
    index traversals; locks cause cross-node synchronization overheads.
Next, even if a \cn fails while performing an index traversal,
    other \cns should be able to make progress.
Finally, the index should support concurrent and consistent updates, allowing
    \dpm threads to perform non-conflicting updates in parallel.
Most state-of-the-art concurrent PM
    indexes satisfy these requirements~\cite{lee2019recipe, Hwang-FAST18, Arulraj-VLDB18, Shaonan-FAST21, Zhangyu-ATC20};  these PM indexes provide
    lock-free reads and log-free in-place writes.
Thus, with such PM indexes, \sysname provides a globally consistent view of data in a scalable manner, independent of the number of \cns.

\vheading{Consistency}.
\sysname guarantees linearizability, the strongest consistency
    level for non-transactional stores~\cite{viotti-CSUR17}.
\sysname ensures that a successful write request commits the data atomically in \dpm, and that subsequent reads return the latest committed value.
To satisfy linearizability, \sysname merges data logs in request order to the metadata index. 
Other core design decisions like \dpart across \cns (\sref{sec:ownership-part}),
    and using indirect pointers for selective replication (\sref{sec:ownership-part}),
    help provide linearizability.
Before reconfiguration or after failure, \sysname merges all pending logs from the \cns involved before allowing the other \cns to serve reads.

\subsection{\cpplain}
\label{sec:adaptive-caching}

It would be prohibitively expensive for \cns to do network round trips (RTs) for every read operation. 
To avoid these overheads, \cns use local DRAM to cache data and metadata.
Because \cns have limited memory,
efficient caching is crucial for high common-case performance.
We introduce \emph{\cpplain} (\cp), a novel caching scheme to efficiently use DRAM at \cn. 

\vheading{Motivation}.
As \dpm is directly accessible to \cns via one-sided RDMA operations with low latency owing to its byte addressability,
\cns can cache not only data in the form of \emph{values} but also metadata in the form of  \emph{shortcuts}.
A \emph{value} entry keeps the entire copy of a \dpm value, so the \cn can access everything locally.
A \emph{shortcut} entry keeps a fixed 64-bit pointer to
  the value in \dpm;
accessing the data incurs a one-sided operation to \dpm.
If neither  value nor shortcut are cached, accessing the value incurs 
  significant overhead: the \cn needs to
traverse a metadata structure in \dpm to find the value's
location and then access the value. Traversing metadata structures
like trees, skip lists, or chaining lists in hash tables,
requires multiple RTs to \dpm or remote
procedures in \dpm, both of which have much higher
overheads than a single one-sided operation~\cite{qingSIGMOD22,aguilera2019designing,zieglerSIGMOD19,pengfeiATC21}.

Caching values improves performance relative to
caching shortcuts, but requires more cache space. This raises
an interesting question: is it better to cache a few values with no overheads upon cache hits, or a larger number of shortcuts with fixed hit overheads? 

The answer is
simple in extreme cases: in highly skewed workloads, where a
small number of hot key-value pairs can fit in the cache,
storing values is better. When workloads are close to uniform distribution with total size larger than the cache, storing shortcuts is better. 
Unfortunately, most workloads fall between these two extremes 
and offer no clear answer.
A simple static caching policy may reserve some fixed ratio
of cache space for storing values and devote the rest to shortcuts.
What should this ratio be? We observe that the efficient ratio is dependent on workload patterns and aggregate memory available for caching.
In a disaggregated system like \sysname that has autoscaling, neither workload patterns nor memory available is known ahead of time, ruling out static policies.

\vheading{Adaptive Policy}.
We introduce \cp, a novel caching policy
that dynamically selects the ratio of values to shortcut entries
as needed. This policy automatically \emph{adapts} to the changes in
workload patterns and to the
changes in the aggregate memory space for caching at \cns due to cluster reconfiguration, as shown in Figure~\ref{fig-dinomo-data-plane}.

\vheading{Insight}.
\cp is based on the following 
insight. 
Performance is highly correlated with number of network RTs, so we seek to minimize that. 
Caching a shortcut reduces RTs from $M$ (where $M$ is the cost of an index lookup) to one,
while caching a value instead of a shortcut reduces RTs from one to zero.
Thus, caching shortcuts provides the bigger gain.
We treat value caching as an optimization
on top of shortcut caching. Value caching is used when we
have spare space in the cache, or when we observe that
storing a value can serve more requests than storing an
equivalent number of shortcuts.
Table~\ref{tab-policy} details the policy.

In \cp, values can be demoted to shortcuts and shortcuts can be evicted. Shortcuts can also be promoted to values. 

\vheading{Demotions}. Demotions occur on cache misses to
make space for a new cache entry.
To demote a value to a shortcut, we pick the least-recently-used key, leveraging temporal locality.
To evict a shortcut, we pick the least-frequently-used key,
in order to preserve frequently used keys in the cache and
cater to skewed workloads.

\begin{table}[t]
  \centering
  \resizebox{.47\textwidth}{!}{%
  \begin{tabularx}{\columnwidth}{lX}
   &  \\ \hline
  \multicolumn{2}{l}{\textbf{\cpplain}} \\ \hline
  BEGIN & We start with an empty cache; start caching values \\
  On a MISS & We cache the shortcut; if we need to make space for the shortcut, we DEMOTE a value (if present) or evict a least frequently used shortcut \\
  On HIT &  We check if we can PROMOTE this shortcut to value; we check if the benefits from caching the value instead of shortcut outweigh the benefits from evicting a suitable number of shortcuts \\
  EVICT & Always evict the least frequently used shortcut \\
  PROMOTE & Promote only if benefits outweigh costs \\
  DEMOTE & Demote if we incur cache misses
  \\ \hline
  \end{tabularx}
  }
  \caption{Summary of the adaptive caching policy}
  \label{tab-policy}
\if\removespace 1
\vspace{-20pt}
\fi
  \end{table}

\vheading{Promotions}. Promotions depend on whether the benefits from caching a
value outweigh the benefits from caching a suitable number
of shortcuts.
To determine if a shortcut $P$ needs to be promoted to a value,
we use the following calculation. If at least $N$ least-frequently-used shortcuts need to be evicted to make space
for caching one value, then the shortcut $P$
needs to satisfy the following relation to be promoted:
\begin{equation}
\begin{aligned}
Hits(P) \times \textrm{Avg. shortcut hit RTs} \geq \\
\sum_{i=1}^{N} Hits(Shortcut_{i}) \times \textrm{Avg. cache miss RTs}
\end{aligned}
\end{equation}
This formula accounts for the two elements of the trade-off:
  the differences in the value and shortcut sizes, and the differences 
  in the cost of a value miss and a shortcut miss.
The left side of the inequality is the number of 
  round-trips saved if we promote shortcut $P$ to a value; the
  right side is the number of additional round-trips
  incurred if we evict $N$ shortcuts to make space for the
  promotion of $P$.
We promote if the savings are greater than the penalty.
Note that the \emph{Avg. shortcut hit RT} is always one, but the
 \emph{Avg. cache miss RT} needs to be determined experimentally,
 which is done by keeping a moving average of past requests.
 

\subsection{\dpartcc}
\label{sec:ownership-part}

When multiple \cns cache the same value,
this can result in consistency overheads (\eg cache invalidation) from ensuring linearizability.
\sysname sidesteps this via \emph{\dpart} (\dip). Owing to DPM architecture where \cns are disaggregated from the shared PM pool, data access and ownership can be independent considerations: it is possible to partition ownership while sharing access to data. This insight motivates \dip that strikes a balance between shared everything and shared nothing.
\dip allows \cns to cache unique data, avoid consistency overheads, and thereby achieve high scalability.
Although similar ideas have been previously used in other contexts~\cite{khattar1999introduction, caulfieldISCA13, snowflake, atulOSDI16},
    we are the first to adapt it for \dpm.

\vheading{Central Idea}. 
\cns have exclusive but temporary ownership of logical, disjoint
  partitions of data.
At any time,
  a partition is accessed
  by only one \cn---its designated owner.
\dip allows \cns to scale without reorganizing data and metadata.

\vheading{Partitioning the ownership}.
Routing nodes maintain the mapping of key ranges to their owner \cns.
Client's requests are routed to the appropriate owner \cn. 
The owner \cn can use its local DRAM to cache data and metadata with high cache locality and provide good read performance.
\sysname does not require cache coherence protocols at \cns,
    as \cns have exclusive access to their partitions.
As scaling \cns increases the total DRAM available
  for caching, \dip scales performance by utilizing the DRAM cache effectively (no redundant copies) and avoiding consistency overheads.

\vheading{Ownership metadata}.
\sysname uses consistent hashing to assign the primary owners for key ranges; \sysname is compatible with other (\eg key-range or hash-based) partitioning algorithms. 
Within a \cn, a key range is further partitioned among its various threads.
Both \cns and \rns maintain the partitioning metadata in a global hash ring, which stores key to \cnfull IP mappings, and a local hash ring, which stores key to thread mappings. 

Whenever the mapping changes, \rns are updated together with \cns. 
Clients cache routing information; when the mapping changes, the \cn they contact will direct them to a routing node to get the latest mapping information.
Each \cn always knows the key range it is supposed to handle, and will refuse requests for other key ranges. 

\vheading{Benefits}. \dpartc provides multiple benefits.

\viheading{High performance}. \sysname achieves high performance in the common case by partitioning the ownership across \cns, allowing multiple \cns to cache unique data partitions with high cache locality. 

\viheading{Scalability}.
By avoiding the overhead for maintaining consistency at \cn caches, \sysname achieves scalability.
    

\viheading{Lightweight reconfiguration}.
\sysname can quickly change the number of \cns without physically reorganizing data or metadata; the current owner empties its cache, completes outstanding operations, hands ownership to the new \cn, and the new owner begins serving requests.
If a \cn fails, partitions owned by the failed \cn can be
assigned to new owners that can immediately serve data.


\vheading{Selective replication}.
Partition-based systems may suffer from load imbalance with highly skewed workloads. 
In these circumstances,
    adding more \cns does not distribute
    the load across available \cns.
Even if the popular key's value is cached in
    a \cn, performance is bottlenecked
    by that \cn's processing or network capacity.
\sysname recognizes such scenarios and 
    shares the ownership of highly popular keys across multiple \cns,
    effectively replicating such keys to
    provide scalability beyond
    a single node's abilities.
The replication metadata is stored along with the mapping information at \rns and \cns while handled similarly. 
Clients cache and use this metadata
    to route requests to primary and secondary owners.

\sysname uses \emph{indirect pointers} to allow \cns to share ownership and read or write the shared key-value pairs consistently.
An indirect pointer points to a location in \dpm that stores a pointer to the value instead of the value itself, and the \cns access the shared value with one-sided CAS operations on the indirect pointers to ensure the linearizable access.
Due to the sharing with indirect pointers, \sysname incurs consistency overheads to balance the load across \cns. \sysname limits these consistency overheads by using indirect pointers only for hot keys.

When a key becomes shared, \sysname installs an indirect pointer to the key in \dpm. 
When a \cn updates a shared key, it writes the value at a new location and atomically updates the indirect pointer. 
A \cn reading a shared key has to first read the indirect pointer and then read the value; thus, shared keys pay a cost that is avoided by default. 
Removing sharing from the key requires the \cns that own the shared key to invalidate it in their caches. Once the invalidation is done, the indirect pointer is removed in \dpm. 

\subsection{Reconfiguration}
\label{sec:control-plane}

\kim{I suggest starting this section with a high-level overview of why/when reconfiguration occurs. We can then point out that we model our policy engine after Anna's and highlight the differences that we use to adapt Anna's policy engine to Dinomo.}

The \mmnode triggers reconfigurations to improve performance when SLOs are violated, to release under-utilized resources, or to tolerate \cn failures.
We first present those policy details and then explain our principled reconfiguration protocol.

\vheading{Policy engine}. 
The policy engine in the \mmnode governs when and what kind of reconfigurations to trigger. 
Our policy engine follows prior autoscaling work~\cite{anna-vldb19}, with simplifications for \sysname; for example, memory consumption is not a consideration in scaling \cns since the memory in a \cn is used as a cache without overflow.
The policy engine allows the configuration of the following parameters: \emph{average/tail latency SLOs}, \emph{over-utilization lower bound} and \emph{under-utilization upper bound}, and \emph{key hotness lower bound} and \emph{key coldness upper bound}.
The \mmnode periodically collects latency information from clients, the average \cn occupancy (\ie CPU working time per monitoring-epoch interval), and the average access frequency for keys from \cns. 
It then proactively detects the latency SLO violations and corrects them dynamically.
Table~\ref{tbl-policy-violations} summarizes reconfiguration scenarios.

\begin{table}[t]
\centering
\resizebox{.47\textwidth}{!}{%
\begin{tabular}{@{}cccc@{}}
\toprule
SLO       & \cn occupancy & Key access freq. & Action           \\ \midrule
Satisfied & Low          & -                & Remove \cn        \\
Violated  & High         & -                & Add new \cn       \\
Violated  & Normal       & High             & Replicate key    \\
Satisfied & Normal       & Low              & De-replicate key \\ \bottomrule
\end{tabular}
}
\caption{Policy violations and \mmnode action}{}
\label{tbl-policy-violations}
\if\removespace 1
\vspace{-20pt}
\fi
\end{table}


\vheading{Cluster membership changes}.
In \sysname, cluster membership is changed under the following scenarios.
First, the \mmnode may detect a \cn failure and notify the alive nodes.
Second, the \mmnode may detect a latency SLO violation (average or tail latency SLO) and find that all the \cns are over-utilized (the minimum occupancy of all \cns is larger than the \emph{over-utilization lower bound}), which triggers the addition of a new \cn.
Third, the \mmnode may detect that there is an under-utilized \cn (its occupancy is lower than \emph{under-utilization upper bound}); if the latency SLOs are not violated, this triggers that \cn's removal.
While ownership mapping is being redistributed due to the membership changes, clients' request latencies can briefly increase.
To prevent the policy engine from over-reacting during the ownership redistribution, \sysname adds or removes at most one node per decision epoch and applies a grace period to allow the system to stabilize before the next decision.

\vheading{Ownership replication changes}.
If the \mmnode detects an SLO violation and notices that all \cns are not over-utilized, then the \mmnode identifies highly popular keys and increases their replication factor.
In detail, the \mmnode considers a key to be highly popular if its average access frequency is greater than the \emph{key hotness lower bound}.
\sysname increases the replication factor $R$ (the number of secondary owners) of a hot key, based on the ratio between the average latency of the hot key and the \emph{average latency SLO}.
The \mmnode considers a key to be cold if its access frequency is below the \emph{key coldness upper bound}.
If the latency SLOs are met and none of the \cns are under-utilized (the \mmnode cannot remove any \cn), the \mmnode identifies cold keys with high replication factors ($R > 1$) and dereplicates them ($R{=}1$).

\vheading{Fault tolerance}.
The \dpm is the source of ground truth in \sysname; it persistently stores data (key-value pairs), metadata (indexing data structures), and other policy information (ownership/replication metadata).
\sysname{}'s \cns and \rns store soft state that can be reconstructed if a node fails.
When a \cn or \rn fails, it retrieves the up-to-date policy information from the \dpm and rebuilds the ownership mapping of key ranges before resuming.
Unlike \rns, a \cn failure changes the ownership mapping among the alive \cns.
The \mmnode ensures that the ownership mapping is corrected before allowing the failed \cn to resume.
After detecting a \cn failure, the \mmnode picks one of the alive \cns to complete the pending operations in the log segments from the failed \cn, and broadcasts the failure to all \sysname components.
On receiving a failure message, \cns and \rns repartition the ownership mapping by updating their local hash rings.

\vheading{Reconfiguration steps}. We now describe how \sysname performs reconfigurations. Broadly, the following steps occur:
\begin{enumerate}[leftmargin=15pt]
\item \cns participating in the reconfiguration are identified (\cns for which the ownership mapping changes)
\item The \cns become unavailable
\item \dpm synchronously merges the data in logs for these \cns
\item The \cns get their new mapping information
\item The \cns become available, and the cluster continues operation
\item The mapping information in the remaining \cns (not participating in the reconfiguration) is updated asynchronously
\item The \rns are asynchronously updated with the new mapping information
\end{enumerate}
The cluster can continue operation at step five because \cns will reject requests for key ranges they do not own. Thus, other \cns can be updated without blocking the nodes undergoing reconfiguration.
In certain special cases, \sysname can perform reconfiguration without blocking any \cns. This can happen when a new partition is being added to \sysname (no previous owner to race with) or when a \cn fails and its partitions are being redistributed.
Note that there is no expensive data copying or movement during reconfiguration. This is the key property that enables lightweight reconfiguration for \sysname.

\subsection{Optimizations}
\label{sec:onesided-async}

\sysname includes optimizations in its data path to reduce CPU bottlenecks and network utilization from \dpm. 

\vheading{One-sided \& asynchronous post processing}.
To minimize the CPU bottlenecks and network utilization, \sysname{}'s data path uses \emph{one-sided operations} with \emph{asynchronous post-processing}. With a one-sided operation (e.g., RDMA reads, writes), a \cn executes directly on the \dpm without involving the \dpm processor. In contrast, two-sided operations (e.g., RPCs) are handled by the \dpm processor. One-sided operations have lower latency and higher bandwidth than two-sided operations~\cite{storm, FaRM-NSDI14,Anuj-ATC16,ChristopherATC13,xingdaOSDI18}, but one-sided operations are limited in functionality~\cite{aguilera2019designing}. 
For the best performance, \sysname uses one-sided operations in the data path and delegates the post-processing of writes to the \dpm processors asynchronously.

\viheading{One-sided reads}.
For reads, an \cn directly returns the value from its cache upon a value hit. On a shortcut hit, it performs a single one-sided operation to retrieve the value in \dpm from the shortcut pointer. On a cache miss, the \cn performs multiple one-sided operations to find the address of the value (index traversals), and uses another one-sided operation to fetch the value from that address.

\viheading{Asynchronous post-processing of writes}.
\sysname batches multiple log entries into a log segment unit and writes them to \dpm using a one-sided RDMA write operation.
With \dip, \sysname can batch the writes for the keys in the same partition without consistency overheads.
The post processing to merge the writes into the metadata index is asynchronously handled by \dpm processors off critical path.
\sysname caches the committed log segments to aid the subsequent reads to be served locally at \cns without expensive network RTs to read the large log segments remotely.
These optimizations have two benefits.
First, it reduces the latency as well as network costs per operation. 
Second, it amortizes the merging operation across all the write operations in a log segment (typically several megabytes in size). 
Because the merging is done asynchronously, the \dpm processors can have lower computing power without significantly affecting \sysname performance.

\section{Implementation}
\label{sec:impl}

We implement \sysname in 10K lines of C++ code.
We use the standard C++ library and several open-source libraries
    including ZeroMQ~\cite{zeromq}, Google Protocol Buffers~\cite{protobuf}, libibverbs~\cite{libibverbs},
    and the PMDK library~\cite{pmdklib}.
This section discusses \sysname{}'s \dpm data structures,
\cp implementation, and cluster management.

\vheading{DPM metadata index}.
\sysname uses RECIPE's P-CLHT (Persistent Cache Line Hash Table)~\cite{lee2019recipe}, which supports lock-free reads and log-free in-place writes, as its metadata index in \dpm.
P-CLHT is a chaining hash table aimed at minimizing the CPU-cache coherence and persistence overheads on PM.
Each bucket in P-CLHT has the size of a single cache line and holds three key-value pairs~\cite{davidASPLOS15}.
The design allows each access/update to the hash table to incur only a single cache-line access/flush in the common case.
For lock-free reads, P-CLHT employs atomic snapshots of key-value pairs.
We modify the index to use RDMA reads for lookups.
On hash collisions, \cns may have to perform multiple one-sided RDMA reads to traverse the hash chain and read the value.
The cacheline-conscious bucket design of P-CLHT, cache-coherent DMA~\cite{FaRM-NSDI14,Anuj-ATC16}, and out-of-place value updates allow us to avoid memory-access races~\cite{ChristopherATC13,singhviSIGCOMM21} between the updates by \dpm processors and one-sided RDMA reads by \cns.

\vheading{DPM log segments}.
\sysname implements 8 MB log segments and handles variable length
    key-value pairs. \cns proactively preallocate log segments for their own use using two-sided operations.
\cns log write operations into \dpm log segments and
    cache them; upon cache misses in \cp,
    \cns have to search cached log segments to find the latest value.
\sysname implements Bloom filters atop cached log segments for
    quick membership queries.
\sysname maintains the following invariant: un-merged log segments are cached in the \cns that wrote them. 
Due to \dip, other \cns won't access these log segments, thus eliminating the need for read operations to check the un-merged log segments on other nodes. 
\cns can add a new log segment to \dpm without blocking until their un-merged log-segment length reaches a certain threshold (default is 2); when the threshold is reached, the critical write paths are blocked until the \dpm processors complete merging below the threshold.
\sysname logs write operations with commit-markers (e.g., a seal byte at the end of the entry~\cite{FaRM-NSDI14,liu-TODS19}) to \dpm log segments
    to ensure crash consistency and to aid recovery.
The DPM index directly points to the values stored in the log entries. 
Since \cns know the address of the log segments they write (and therefore where values are stored), they can produce and locally cache shortcuts to values in \dpm without an extra round trip. 
To garbage collect stale log segments, \sysname maintains per-log-segment counters that reflect the number of valid and invalid values in each log segment.
Once the number of invalid values matches the total number of values in a log segment, a \dpm processor garbage collects the log segment.


\vheading{DPM persistence}.
While merging log segments, \sysname{}'s \dpm processing threads persist all the writes to the \dpm index structure using \texttt{CLWB}, \texttt{sfence}, and non-temporal store instructions~\cite{rudoff2017persistent}.
RDMA currently does not support durable RDMA writes. 
However, the proposed durable write in the IETF standards working document~\cite{Talpey2016RDMADW} behaves similar to a non-durable write, requiring one network round trip. 
Our implementation currently uses non-durable writes, and we plan to update these to durable writes once they become available~\cite{Kim-SIGCOMM18}. 


\vheading{\cp}.
\cp is implemented using standard C++ libraries. \cp uses two unordered maps
    to store values and shortcuts. 
    Least recently used values  
    and least frequently used shortcuts are evicted. 
    The key access frequency
    is tracked using a multimap.
The shortcut entries in \cp contain a pointer to a \dpm value, and the \dpm value length.
The value entries have two more extra fields, an access count and a copy of the \dpm value.
In \cp, demoted values are cached as shortcuts,
    and shortcuts being promoted inherit their access counts to preserve their access history.

\vheading{Cluster management}.
\sysname uses Kubernetes~\cite{kubernetes} for cluster orchestration.
Pods are the smallest deployable units in Kubernetes.
Each \sysname component is instantiated in a separate Kubernetes pod with a corresponding Docker~\cite{docker} container.
\sysname uses Kubernetes to add/remove \cn pods and restart failed pods.
The \mmnode pod is colocated with the Kubernetes master.
The \mmnode's policy engine adds/removes \cn pods by running simple bash scripts executing \texttt{kubectl}~\cite{kubectl} commands to the Kubernetes master.
The Kubernetes master keeps track of pod status using heartbeats, and the \mmnode uses this information to detect failures in \cn pods.
    

\section{Evaluation}
\label{sec:evaluation}

We evaluate the performance of \sysname and
    study the breakdown of the benefits from \dpartcc (\dip),
    \cpplain (\cp), and selective replication. We design our experiments
    to answer the following questions:
\begin{itemize}[leftmargin=15pt]
    \item Does \cp help reduce network round trips? How does it fare against other caching policies? 

    \item How much impact does the compute capacity in \dpm have on the overall throughput in \sysname?

    \item How does \sysname fare performance and scalability against the state-of-the-art?

    \item What fraction of \sysname's benefits can be attributed to the \dip architecture and the \cp caching?
    
    \item How elastic and responsive is \sysname while handling bursty workloads, load imbalance, and \cn failures?
\end{itemize}

\vheading{Comparison points}.
As our baseline, we use Clover~\cite{tsai2020disaggre}, a state-of-the-art and open-source key-value store designed for \dpm.
Clover has a shared-everything architecture with
    a shortcut-only cache at its \cns.
\cns perform out-of-place
    updates to the data in \dpm, and incur additional overheads
    to provide strong consistency. For example, stale cached
    entries require \cns to walk through a chain of versions to find the most recent data in \dpm. 

Besides Clover, we compare \sysname with two variants, \sysname-S and \sysname-N.
\sysname uses three techniques: \cp, \dip, and selective replication. 
\sysname-S employs shortcut-only caching; it is otherwise identical to \sysname.
As the source code of AsymNVM~\cite{ma2020asymnvm} is not publicly available, we implement \sysname-N to compare \sysname with a shared-nothing counterpart; it uses \cp but partitions data and metadata in \dpm, where each partition is exclusively accessed by a single \cn without selective replication.

Comparing \sysname-S with Clover highlights the benefits of partitioning ownership in \dip, and comparing \sysname with \sysname-S shows the benefits from \cp.
We also investigate the trade-off from sharing data in \dip by comparing \sysname with \sysname-N.

\vheading{Experiment setup}.
We use Kubernetes pods to represent all of the node instances in the \sysname cluster.
We restrict the host resources assigned to the pods depending on the node types' features to emulate the asymmetric \dpm architecture (i.e., \cns have more-capable computation but smaller memory than \dpm). 
Each individual pod is pinned to a separate server for resource isolation purposes.

We deploy \sysname on the Chameleon Cloud~\cite{chameleonATC20}, an experimental large-scale testbed for cloud research.
We use InfiniBand-enabled (IB-enabled) servers as hosts for \cns and the \dpm; each two-socket server has Intel Xeon E5-2670v3 processors, 24 cores at 2.30 GHz in total, and 128 GB DRAM.
The shared \dpm uses a maximum of 4 threads and 110 GB of DRAM as a proxy for the PM, which is registered to be RDMA-accessible.
Each \cn uses a maximum of 8 threads and 1 GB of DRAM for caching (${\approx}1$\% of the \dpm size).
\dpm and the \cns are connected by Mellanox FDR ConnectX-3 adapters with 56 Gbps per port.
We emulate PM using DRAM as  performance is constrained by the network rather than PM or DRAM: network latency (1--20 us) is at least 10\myx higher than DRAM or PM latencies (100s of ns); network bandwidth (7GB/s) is lower than PM bandwidth (32GB/s Read / 11.2GB/s Write)~\cite{anujSOCC20, assise-osdi20}.

The external servers that run application workloads, henceforth termed \emph{client nodes}, and the routing service do not need a high-speed interconnect with the \cns or \dpm.
Hence, for client nodes and routing nodes (\rns), we use two-socket servers with AMD EPYC 7763 processors, 128 cores at 2.45 GHz in total, 256 GB of DRAM, and a 10 Gbps Ethernet NIC.
Each client node uses 64 threads to run a closed-loop workload with one or more outstanding requests per thread.
We use a single \rn with 64 threads. The same routing layer is used across all KVS variants in our evaluation.
Apart from the data plane components (\cns and \dpm), \sysname, \sysname-N, and \sysname-S use a specific control-plane instance for the \mmnode, which is deployed on a server (same server configuration as the \rns) with a single thread.
For Clover, we use an extra IB-enabled server (same server configuration of the \cns) for its metadata server with 6 threads (4 workers, 1 epoch thread, 1 GC thread).

\begin{figure}[!t]
  \begin{center}
  \includegraphics[width=0.47\textwidth]{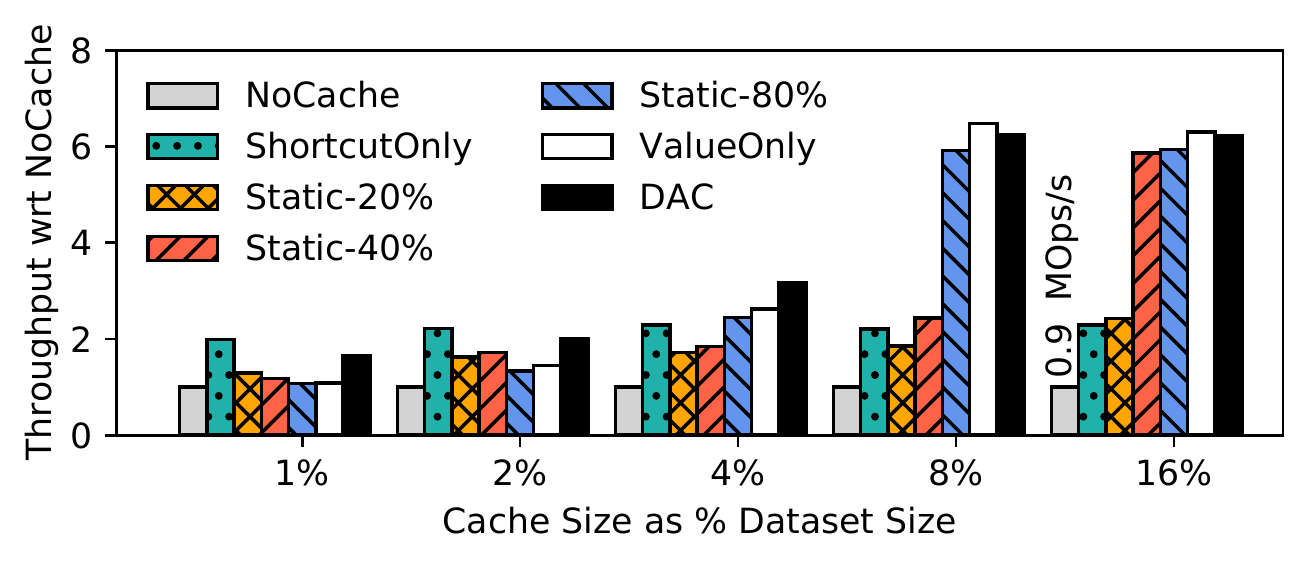}
  \end{center}
\if\removespace 1
  \vspace{-15pt}
\fi
  \caption{Performance comparison of cache policies}{}
  \label{fig-eval-micro-cache}
  \if\removespace 1
  \vspace{-10pt}
  \fi
\end{figure}
\begin{table}[!t]
\small
\centering
\begin{tabular}{@{}c|p{0.3cm}cccccc@{}}
\toprule
Cache size as & No & Shortcut & Static & Static & Static & Value & \multirow{2}{*}{\textbf{DAC}} \\
\% dataset size & Cache & Only & 20\% & 40\% & 80\% & Only &  \\ \midrule
1\% & 5.0 & 1.5 & 2.0 & 2.5 & 3.6 & 4.2 & \textbf{1.4} \\
2\% & 5.0 & 1.1 & 1.0 & 1.0 & 2.2 & 3.1 & \textbf{0.9} \\
4\% & 5.0 & 1.1 & 1.0 & 0.8 & 0.5 & 1.4 & \textbf{0.4} \\
8\% & 5.0 & 1.1 & 0.8 & 0.5 & 0.1 & 0.1 & \textbf{0.1} \\
16\% & 5.0 & 1.1 & 0.5 & 0.1 & 0.1 & 0.1 & \textbf{0.1} \\ \bottomrule
\end{tabular}
\mycaption{RTs/operation}{\cp has lowest RTs/op across different static caching strategies in all settings.\vijay{Table 4 doesn't add much to the paper. we can remove if we need space.}}
\label{tab-micro-cache-rtts}
\if\removespace 1
\vspace{-20pt}
\fi
\end{table}

\vheading{Workloads and configurations}.
We use YCSB-style workloads~\cite{anna-vldb19,cooperSOCC10} with five request patterns: read-only (100\% reads), read-mostly (95\% reads/5\% updates and 95\% reads/5\% inserts), and write-heavy (50\% reads/50\% updates and 50\% reads/50\% inserts).
These workloads use 8B keys and 1KB values and the following Zipfian coefficients: 0.99 (the YCSB-default value) for moderate skew, 2 for high skew, and 0.5 for low skew (close to uniform).
For each experiment, we first load 32 GB of data (key-value pairs) and then write up to 100GB of data during the workload including inserts.
With 16 \cns, each equipped with a 1GB cache, the \cns can cache up to 50\% of the loaded dataset.
We generate the workload from the client nodes and measure system throughput and other profiling metrics, averaging them over a 10-second interval.
    
\subsection{Microbenchmark}\label{sec:eval-micro}
We use micro-benchmarks to investigate several issues.
We first consider whether DAC is an effective caching strategy.
Next, we explore how much compute capacity the \dpm requires to prevent the asynchronous merging of writes from becoming the bottleneck; based on the results, we also discuss how our use of DRAM to emulate PM affects our results.


\vheading{\cp}.
The \cn caches can be used to store values, shortcut pointers, or a mix of both. 
To evaluate \cp against different caching strategies, we use a single \cn with 16 threads. We first load 30M key-value pairs into \sysname with 8B keys and 64B values. 
We then run a read-only workload with a working set of uniformly-distributed 1.5M keys (5\% of the dataset) to evaluate performance.
We generate the workload locally and measure the peak throughput within the \cn by varying the available DRAM for caching from 1\%--16\% of the dataset size.
We configure \sysname to use different caching policies (Figure~\ref{fig-eval-micro-cache}).
The static-X policies reserve X\% of their cache size for storing values; the rest of the cache is used for shortcuts.
All non-\cp policies use LRU to evict entries.

Figure~\ref{fig-eval-micro-cache} shows the read throughput obtained with the different cache policies. With an aggregate cache size of 2\% of the dataset, a shortcut-only cache performs best, whereas with a cache size 4\myx as large, a value-only cache performs best. The aggregate cache size is dependent on the number of active \cns, which may dynamically change with cluster reconfiguration or \cn failures.
Therefore, a static caching policy is not a good fit. The right policy depends upon the workload patterns and aggregate cache size. 

Despite not knowing the workload patterns or the aggregate cache size, \cp is within 
16\% of the best performing policy, in all settings. 
With a medium-sized cache that is 4\% of the dataset size, \cp  exceeds the performance of both shortcut-only and value-only caching policies by taking advantage of both.
Further, as shown in Table~\ref{tab-micro-cache-rtts}, \cp has the lowest number of round trips per operation compared to all static policies, reducing 
the network utilization and providing high performance.

\begin{figure}[!t]
  \begin{center}
  \includegraphics[width=0.47\textwidth]{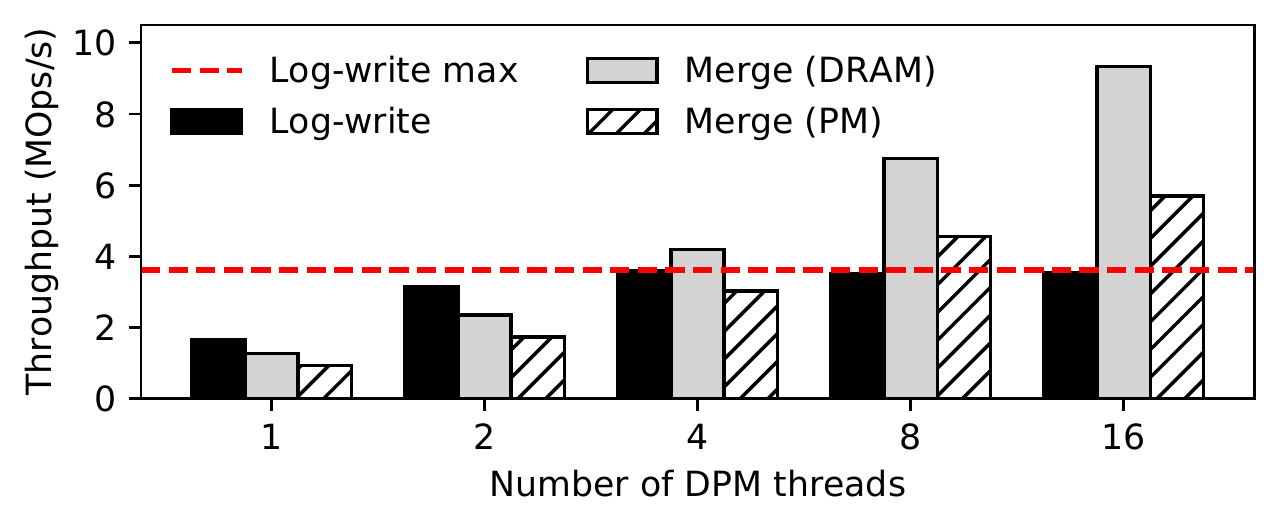}
  \end{center}
\if\removespace 1
  \vspace{-15pt}
\fi
  \mycaption{Performance impact of \dpm compute capacity}{The log-write throughput approaches the max with 4 threads on DRAM, while requiring more threads on PM.}
  \label{fig-eval-micro-app}
  \if\removespace 1
  \vspace{-10pt}
  \fi
\end{figure}
\begin{figure*}
  \begin{center}
    \includegraphics[width=1.0\textwidth]{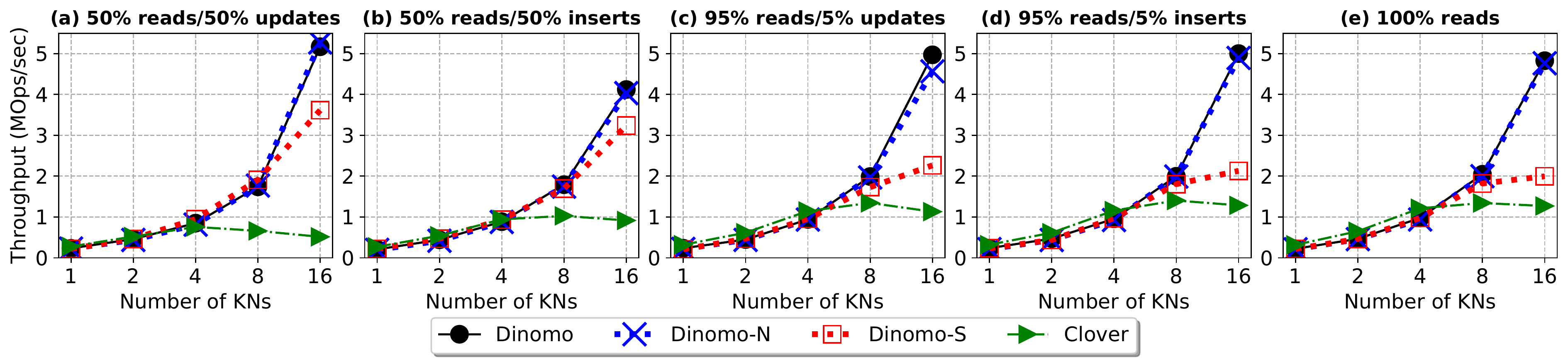}
  \end{center}
\if\removespace 1
  \vspace{-10pt}
\fi
  \mycaption{Performance scalability}{\sysname-S vs. Clover highlights the benefits from \dip. \sysname vs. \sysname-S highlights the benefits from \cp. \sysname vs. \sysname-N shows the performance trade-offs between sharing data and \dip.} 
  \label{fig-perf}
\end{figure*}

\begin{table*}[t]
\centering
\resizebox{\textwidth}{!}{%
\begin{tabular}{@{}p{0.15cm}c|cccp{0.2cm}p{0.2cm}p{0.3cm}|cccp{0.2cm}p{0.2cm}p{0.3cm}|cccp{0.2cm}p{0.2cm}p{0.3cm}|cccp{0.2cm}p{0.2cm}p{0.3cm}|cccp{0.2cm}p{0.2cm}l@{}}
\cmidrule(l){3-32}
\multicolumn{2}{c|}{\multirow{2}{*}{}} & \multicolumn{6}{c|}{\textbf{50\% reads/50\% updates}} & \multicolumn{6}{c|}{\textbf{50\% reads/50\% inserts}} & \multicolumn{6}{c|}{\textbf{95\% reads/5\% updates}} & \multicolumn{6}{c|}{\textbf{95\% reads/5\% inserts}} & \multicolumn{6}{c}{\textbf{100\% reads}} \\ \cmidrule(l){3-32} 
\multicolumn{2}{c|}{} & \multicolumn{3}{c|}{Cache Hit Ratio} & \multicolumn{3}{c|}{RTs/op} & \multicolumn{3}{c|}{Cache Hit Ratio} & \multicolumn{3}{c|}{RTs/op} & \multicolumn{3}{c|}{Cache Hit Ratio} & \multicolumn{3}{c|}{RTs/op} & \multicolumn{3}{c|}{Cache Hit Ratio} & \multicolumn{3}{c|}{RTs/op} & \multicolumn{3}{c|}{Cache Hit Ratio} & \multicolumn{3}{c}{RTs/op} \\ \midrule
\multicolumn{2}{c|}{KVSs} & \textbf{D} & DS & \multicolumn{1}{c|}{C} & \textbf{D} & DS & C & \textbf{D} & DS & \multicolumn{1}{c|}{C} & \textbf{D} & DS & C & \textbf{D} & DS & \multicolumn{1}{c|}{C} & \textbf{D} & DS & C & \textbf{D} & DS & \multicolumn{1}{c|}{C} & \textbf{D} & DS & C & \textbf{D} & DS & \multicolumn{1}{c|}{C} & \textbf{D} & DS & C \\ \midrule
\multicolumn{1}{c|}{} & 1 & \textbf{\begin{tabular}[c]{@{}c@{}}100\\ (52)\end{tabular}} & 100 & \multicolumn{1}{c|}{100} & \textbf{0.2} & 0.5 & 2.8 & \textbf{\begin{tabular}[c]{@{}c@{}}100\\ (53)\end{tabular}} & 100 & \multicolumn{1}{c|}{100} & \textbf{0.3} & 0.5 & 2.5 & \textbf{\begin{tabular}[c]{@{}c@{}}100\\ (52)\end{tabular}} & 100 & \multicolumn{1}{c|}{100} & \textbf{0.5} & 1.0 & 2.1 & \textbf{\begin{tabular}[c]{@{}c@{}}100\\ (53)\end{tabular}} & 100 & \multicolumn{1}{c|}{100} & \textbf{0.4} & 0.9 & 2.1 & \textbf{\begin{tabular}[c]{@{}c@{}}100\\ (52)\end{tabular}} & 100 & \multicolumn{1}{c|}{100} & \textbf{0.5} & 1.0 & 2.0 \\ \cmidrule(l){2-32} 
\multicolumn{1}{c|}{\multirow{5}{*}{\rotatebox[origin=c]{90}{\# KNs}}} & 2 & \textbf{\begin{tabular}[c]{@{}c@{}}100\\ (55)\end{tabular}} & 100 & \multicolumn{1}{c|}{95} & \textbf{0.2} & 0.5 & 3.5 & \textbf{\begin{tabular}[c]{@{}c@{}}100\\ (55)\end{tabular}} & 100 & \multicolumn{1}{c|}{95} & \textbf{0.2} & 0.5 & 2.5 & \textbf{\begin{tabular}[c]{@{}c@{}}100\\ (56)\end{tabular}} & 100 & \multicolumn{1}{c|}{96} & \textbf{0.4} & 0.9 & 2.3 & \textbf{\begin{tabular}[c]{@{}c@{}}100\\ (55)\end{tabular}} & 100 & \multicolumn{1}{c|}{95} & \textbf{0.4} & 1.0 & 2.1 & \textbf{\begin{tabular}[c]{@{}c@{}}100\\ (55)\end{tabular}} & 100 & \multicolumn{1}{c|}{96} & \textbf{0.4} & 1.0 & 2.0 \\ \cmidrule(l){2-32} 
\multicolumn{1}{c|}{} & 4 & \textbf{\begin{tabular}[c]{@{}c@{}}100\\ (59)\end{tabular}} & 100 & \multicolumn{1}{c|}{88} & \textbf{0.2} & 0.5 & 4.2 & \textbf{\begin{tabular}[c]{@{}c@{}}100\\ (59)\end{tabular}} & 100 & \multicolumn{1}{c|}{88} & \textbf{0.3} & 0.5 & 2.6 & \textbf{\begin{tabular}[c]{@{}c@{}}100\\ (59)\end{tabular}} & 100 & \multicolumn{1}{c|}{89} & \textbf{0.4} & 0.9 & 2.5 & \textbf{\begin{tabular}[c]{@{}c@{}}100\\ (59)\end{tabular}} & 100 & \multicolumn{1}{c|}{89} & \textbf{0.4} & 0.9 & 2.2 & \textbf{\begin{tabular}[c]{@{}c@{}}100\\ (59)\end{tabular}} & 100 & \multicolumn{1}{c|}{90} & \textbf{0.4} & 1.0 & 2.1 \\ \cmidrule(l){2-32} 
\multicolumn{1}{c|}{} & 8 & \textbf{\begin{tabular}[c]{@{}c@{}}100\\ (68)\end{tabular}} & 100 & \multicolumn{1}{c|}{82} & \textbf{0.2} & 0.4 & 5.1 & \textbf{\begin{tabular}[c]{@{}c@{}}100\\ (66)\end{tabular}} & 100 & \multicolumn{1}{c|}{83} & \textbf{0.3} & 0.5 & 2.6 & \textbf{\begin{tabular}[c]{@{}c@{}}100\\ (68)\end{tabular}} & 100 & \multicolumn{1}{c|}{85} & \textbf{0.3} & 0.9 & 2.6 & \textbf{\begin{tabular}[c]{@{}c@{}}100\\ (68)\end{tabular}} & 100 & \multicolumn{1}{c|}{84} & \textbf{0.3} & 0.9 & 2.3 & \textbf{\begin{tabular}[c]{@{}c@{}}100\\ (68)\end{tabular}} & 100 & \multicolumn{1}{c|}{84} & \textbf{0.3} & 1.0 & 2.3 \\ \cmidrule(l){2-32} 
\multicolumn{1}{c|}{} & 16 & \textbf{\begin{tabular}[c]{@{}c@{}}100\\ (77)\end{tabular}} & 100 & \multicolumn{1}{c|}{77} & \textbf{0.1} & 0.5 & 8.7 & \textbf{\begin{tabular}[c]{@{}c@{}}100\\ (81)\end{tabular}} & 100 & \multicolumn{1}{c|}{77} & \textbf{0.1} & 0.5 & 2.7 & \textbf{\begin{tabular}[c]{@{}c@{}}100\\ (83)\end{tabular}} & 100 & \multicolumn{1}{c|}{79} & \textbf{0.1} & 0.9 & 2.8 & \textbf{\begin{tabular}[c]{@{}c@{}}100\\ (88)\end{tabular}} & 100 & \multicolumn{1}{c|}{80} & \textbf{0.1} & 0.9 & 2.4 & \textbf{\begin{tabular}[c]{@{}c@{}}100\\ (87)\end{tabular}} & 100 & \multicolumn{1}{c|}{79} & \textbf{0.1} & 1.0 & 2.4 \\ \bottomrule
\end{tabular}%
}
  \mycaption{Profiling numbers}{For \sysname (D), \sysname-S (DS), and Clover (C), cache hit ratios and RTs per operation are measured across all \cns. For \sysname, we report both the total cache hit \% and the \% due to values (in parentheses). We omit the profiling numbers for \sysname-N since it has similar trends as \sysname.\vijay{table doesn't add much to the paper IMO. we can just add text explaining what we saw in the profiling numbers in the appropriate place in paper}}
\label{tbl-perf-profile}
\if\removespace 1
\vspace{-20pt}
\fi
\end{table*}

\vheading{Asynchronous post processing}.
A delay in merging log segments due to the limited compute capacity in \dpm can block the critical path of \cns writing logs.
To evaluate this impact from the worst-case scenario in our setup, we run an insert-only workload using 16 \cns and 8 client nodes; this is the most compute-intensive workload, as it incurs structural changes to the \dpm index (\eg resizing hash table).
We first load 32GB of data and then run the workload writing up to 100GB of data into \dpm with 8B keys and 1KB values.
We measure the peak throughput of log writing and merging for different \dpm thread counts.
For the log-write throughput, we collect the aggregate throughput across 16 \cns every 10 seconds for 30 seconds and average them; the log-write max is the maximum throughput the \cns can obtain if they never wait for \dpm to merge logs.
To measure the merge throughput, we pre-generate log segments locally on \dpm for the dataset and then measure the performance of merging those log segments.
As our testbed has no IB-enabled PM machines, we measure the merge throughput on PM using a local PM machine (Intel Xeon Silver 4314 CPU with 16 cores at 2.4GHz and 512GB Intel Optane DC PM on 4 NVDIMMs) to estimate the impact from using PM for \dpm.


We make a number of observations based on the results in Figure~\ref{fig-eval-micro-app}.
First, we observe that to write logs at the maximum rate, \dpm should have enough computing capability to merge at the log-write max rate; four or more \dpm threads are required from our setup.
Second, we observe that because of PM's higher access latency, PM merge throughput is lower than DRAM; when using four threads, the lower PM merge throughput can become the bottleneck.
Third, we confirm despite in-DIMM write amplifications~\cite{yang20-pm, anujSOCC20}, merge operations consume PM write bandwidth only up to 2GB/s (monitored by PCM~\cite{pcm}); 9.2GB/s out of the maximum (11.2GB/s) is still available to absorb incoming writes from the KNs over the network, making the network (7GB/s) the bottleneck rather than PM.

We conclude that, in some scenarios, using PM instead of DRAM requires a higher number of \dpm threads to prevent the merging delay from becoming the bottleneck. 
However, even in this worst-case scenario, PM merge throughput with 4 threads was only 16\% lower than log-write max;
for more realistic scenarios with a mix of read and write operations (as used in our following end-to-end experiments), \dpm should be able to operate with the same number of threads (4 threads or more) on both PM and DRAM for 16 \cns. 




\subsection{Performance and Scalability}
\label{sec:eval-perf-scale}

We now compare the end-to-end performance and scalability of \sysname, \sysname-S (\sysname with a shortcut-only cache), \sysname-N (\sysname with \cp and data/metadata partitioning), and Clover.
We use workloads with moderate skew (Zipf 0.99) to observe the performance and scalability in the common case.
We use 8 client nodes to run these workloads and measure the peak throughput by increasing the outstanding requests per client thread until the \cns' CPUs are saturated.
After a 1-minute warm-up period, we collect the aggregate throughput across \cns every 10 seconds for 40 seconds and average them.
In this experiment, the number of \cns is fixed, and hence there is no reconfiguration.
However, the overhead to monitor system statistics (which are  used to trigger reconfiguration) is reflected in the measurement of \sysname and its variants.
We profile the workload and collect metrics such as aggregate cache hit ratio and the average number of network round trips per operation (RTs/op) 
across all \cns, as shown in Table~\ref{tbl-perf-profile}.
    
As shown in Figure~\ref{fig-perf}, \sysname's throughput scales to 16 \cns.
In contrast, Clover's throughput does not scale beyond 4 \cns due to either a network bottleneck or the CPU bottleneck from its metadata server.
With 16 \cns, \sysname outperforms Clover by at least 3.8\myx across all workloads. 
\sysname-S does not scale beyond 8 \cns in read-dominated workloads  because of network bottlenecks.
The performance of \sysname and \sysname-N is almost on par
(max difference is 11\%).
We observe that both \sysname and \sysname-N achieve high performance due to high cache locality at \cns resulting from partitioning.
While partitioning data and metadata in \sysname-N also reduces synchronization overheads, we did not notice significant benefit due to this in the tested workloads. 

\vheading{\dip enables scalable performance}.
We observe that increasing the number of \cns from 1 to 16 reduces the cache hit ratio in Clover across all workloads (Table~\ref{tbl-perf-profile}).
This performance drop is counterintuitive, as the DRAM available for caching
increases with the number of \cns.
However, in shared-everything architectures \cns can handle any request, so multiple \cns may incur cache misses on the same key.
With more \cns, even with moderate skew, the redundant cache misses increase.
In summary, shared-everything architectures do not provide good cache locality and prevent the efficient use of \cn-side memory for caching.
In contrast, \dip partitions the ownership of keys across \cns, providing high cache locality for requests and eliminating redundant shortcuts at multiple \cns.
Note that, for these workloads, \sysname-S sees a 100\% hit ratio across all \cns and with any number of \cns (Table~\ref{tbl-perf-profile}).

\vheading{\cp boosts performance and scalability}.
\sysname has a higher cache hit rate (from values) with more \cns
and takes fewer RTs/op, compared to both \sysname-S and Clover.
\sysname-S has higher network costs: up to 10\myx more RTs/op than \sysname. Clover is even worse: from 4\myx to 87\myx more RTs/op than \sysname, due to shortcut-only caching and a lack of locality that results in consistency overheads and redundant caching.
The aggregate memory available for caching increases with \cns for all systems.
However, \cp helps \cns cache more values (as opposed to shortcuts), and thus incur fewer round trips to \dpm per operation.
In \sysname, the cache hit \% from values increases from 52\% with 1 \cn up to 88\% with 16 \cns across all workloads (Table~\ref{tbl-perf-profile}).
With 1 \cn, \sysname caches more shortcuts, incurring 1 RT at a cache hit, while with 16 \cns, \sysname caches more values, and hence takes fewer RTs/op (0.1 RTs/op across all workloads).
\sysname has fewer RTs/op in write-heavy workloads on average than read-dominated workloads, as \cns persists multiple write operations in a batch with 1 RT on \dpm.
Overall, we see that \cp is effective in reducing RTs to \dpm.

\subsection{Elasticity}
\label{sec:eval-elasticity}
We now demonstrate \sysname can elastically scale the number of \cns, balance loads across \cns, and tolerate failures. We use a workload with 50\% reads and
50\% updates with three different skew distributions.
When a reconfiguration is triggered in this workload, any pending writes must be merged to \dpm before the reconfiguration can proceed.
We run a client node with one outstanding request per thread a time.

\vheading{Policy Variables}.
We set the parameters of the policy engine (\sref{sec:control-plane}) and design the experiments to trigger various forms of reconfiguration. 
We use an \emph{average latency SLO} of 1.2ms 
and a \emph{tail latency SLO} (99-percentile latency) of 16ms. 
The \emph{over-utilization lower bound} is configured to be 20\% \cn occupancy, and the \emph{under-utilization upper bound} is set to 10\% \cn occupancy. 
Furthermore, we configure the \emph{key-hotness lower bound} to 3 standard deviations above the mean key access frequency and the \emph{key-coldness upper bound} to 1 standard deviation below the mean. 
Note that the goal of the experiments is to study the elasticity of \sysname 
under various scenarios; we chose these policy parameters as simple triggers for these scenarios, not as an indication of the best policies.


\begin{figure}
  \begin{center}
    \includegraphics[width=0.47\textwidth]{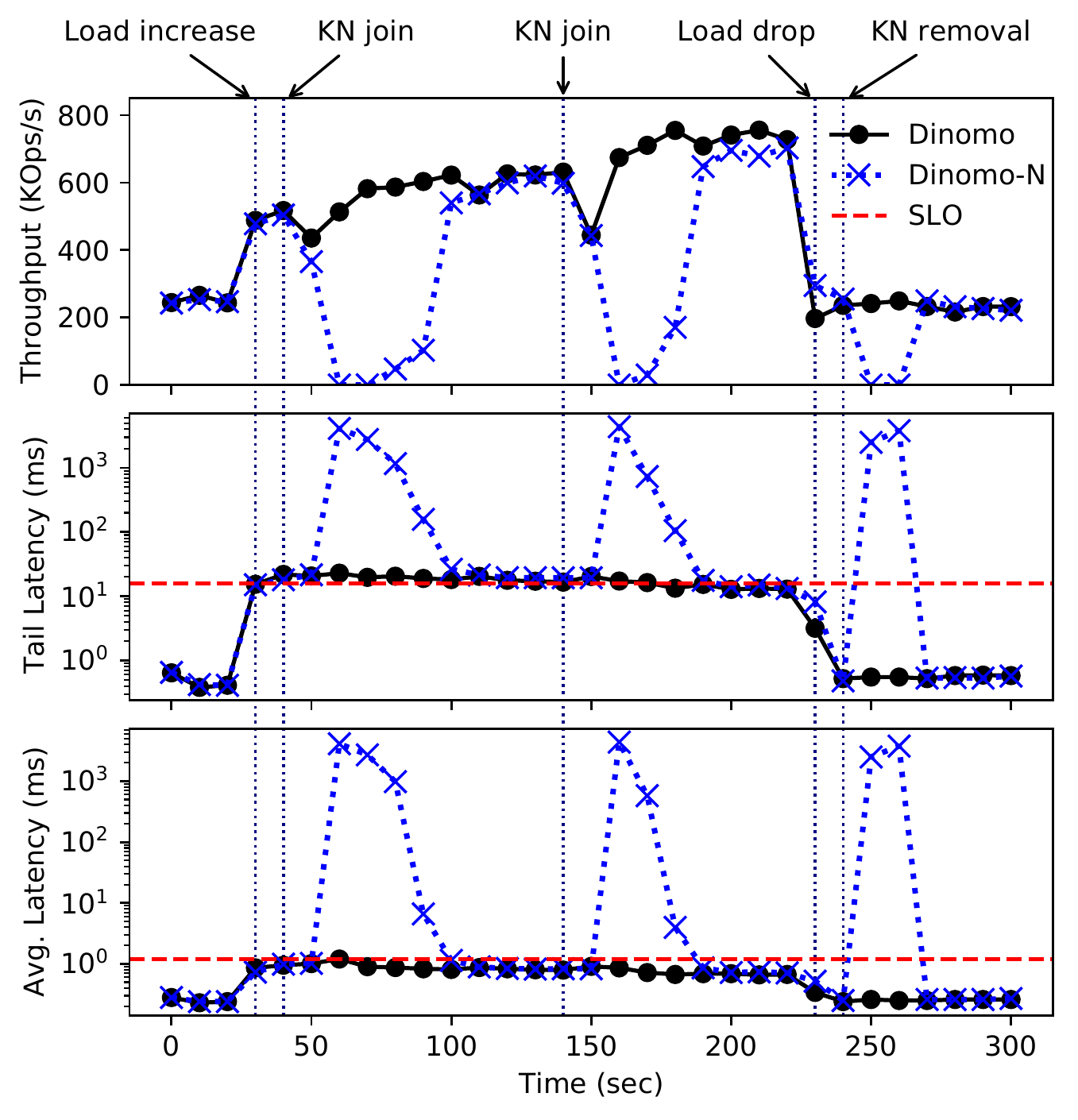}
  \end{center}
\if\removespace 1
  \vspace{-15pt}
\fi
  \caption{Latency and throughput of \sysname and \sysname-N over time while changing the load and number of \cns}{}
  \label{fig-scale-out}
  \if\removespace 1
  \vspace{-10pt}
  \fi
\end{figure}

\vheading{Auto Scaling}.
We evaluate \sysname with bursty, irregular workloads and compare its elasticity in scaling \cns with \sysname-N.
We were unable to run Clover for this experiment because Clover has no implementation for auto-scaling \cns.
We produce scenarios where a new \cn is required or an existing \cn is no longer needed. 
Recall that \sysname adds new \cns automatically only if a latency SLO is violated, the \cns are over-utilized, and an additional \cn is available. 
\sysname automatically evicts a \cn only if the latency SLOs are met and the \cn is underutilized. 
The grace period after each reconfiguration is configured to 90 seconds.

To produce a bursty workload, we start running the workload with low skew (Zipf 0.5) on \sysname using 1 client node for 20 seconds. 
We then increase the load on \sysname by 7\myx by adding 7 additional client nodes. 
We observe the performance of \sysname for a few minutes until it stabilizes, and at the 230-second mark, we remove 7 client nodes to lower the load by 7\myx again. 
Figure~\ref{fig-scale-out} shows the behavior of \sysname and \sysname-N during this experiment.

\sysname and \sysname-N meet the latency SLOs until the load increases at 30 seconds, when the \mmnode detects a latency SLO violation: the tail latency SLO
is exceeded. The \mmnode then observes that \cns are over-utilized (minimum \cn occupancy in \sysname is about 35\%), and hence corrects the situation by adding a new \cn. Once the new \cn comes online at 40-50 seconds, \sysname shows a brief latency increase and throughput dip, as the nodes update their hash rings. However, \sysname-N experiences a 40-second latency spike and throughput dip at 60 seconds, where the throughput drops to 0 due to the processing delay during data reorganization. After a 90-second grace window, although the average latency SLO is met, the tail latency SLO is still violated. \sysname and
\sysname-N react to the situation by adding another \cn. Again, \sysname only
sees a brief increase in latency, while \sysname-N's latency increases for 30 seconds. After the grace window, as both latency SLOs are met, \sysname and \sysname-N do not take any further actions.

At 230 seconds, the load is suddenly reduced. In the next 10 seconds, the \mmnode detects an under-utilized \cn with less than 10\% occupancy. As the latency SLOs are met, the policy engine triggers the \cn eviction. While removing the under-utilized \cn, \sysname sees a brief rise in average and tail latency without violating SLOs. However, \sysname-N shows a 20-second throughput
dip and latency spike before stabilizing.

Overall, we see that \sysname is more responsive with fewer throughput and latency disruptions than \sysname-N and can automatically scale \cns as required by changes in load.

\begin{figure}
  \begin{center}
    \includegraphics[width=0.47\textwidth]{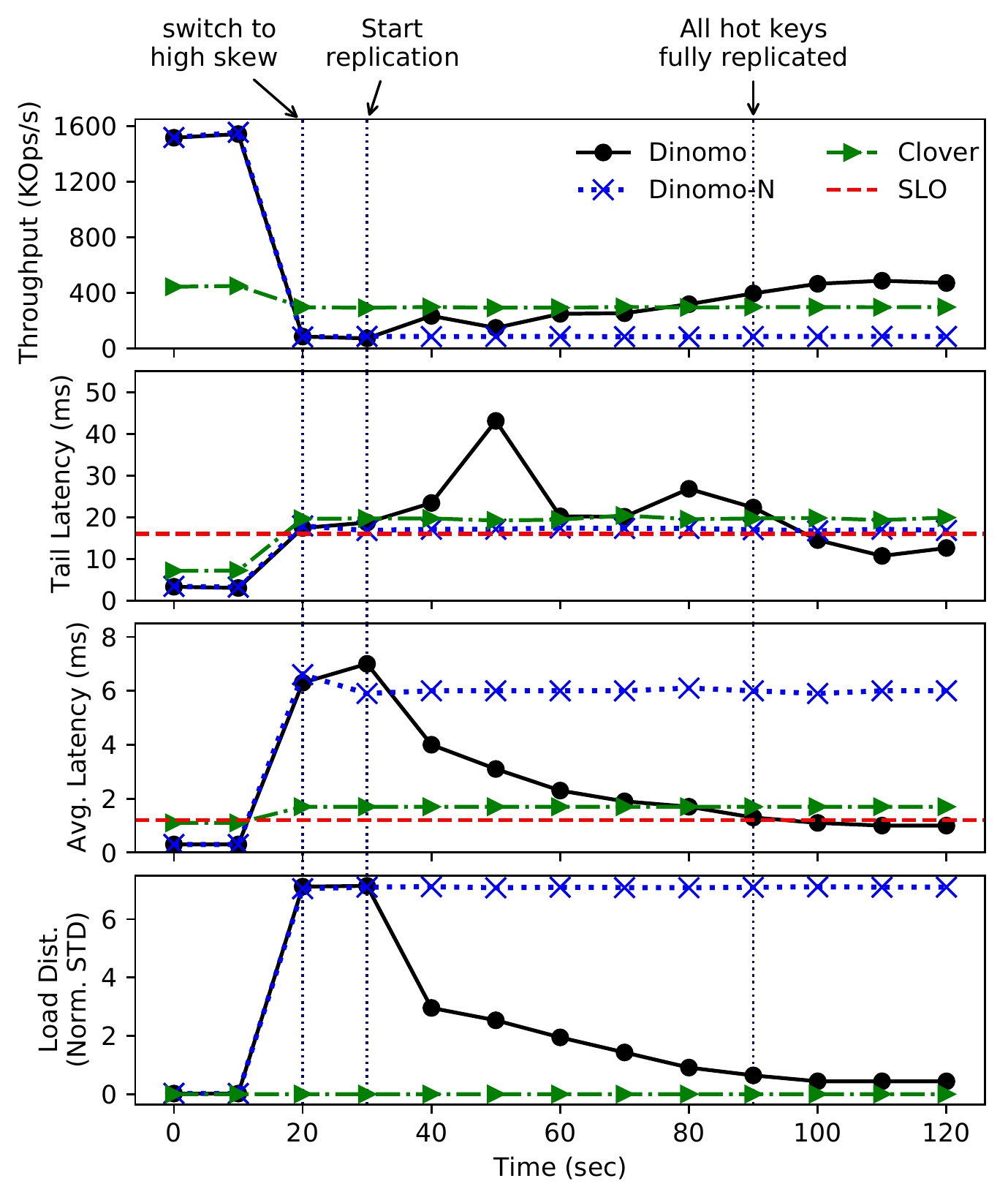}
  \end{center}
\if\removespace 1
  \vspace{-15pt}
\fi
  \caption{Latency and throughput of \sysname, \sysname-N, and Clover over time while running a highly-skewed workload}{}
  \label{fig-load-balance}
  \if\removespace 1
  \vspace{-10pt}
  \fi
\end{figure}

\vheading{Load Balancing}.
We now describe how \sysname handles non-uniform load on its \cns and scales its throughput for hot spots, in comparison to \sysname-N and Clover.
To handle these scenarios, recall that \sysname uses selective replication; this mechanism is triggered only if a latency SLO is violated due to a few hot keys and the \cns are not over-utilized.

For these experiments, we use a skewed workload with 8 client nodes and 16 \cns.
We start the experiments with a low-skew workload (Zipf 0.5) and then switch to the highly-skewed workload (Zipf 2). 
\sysname{}'s policy engine checks that the \cns are not over-utilized (minimum occupancy lower than 10\%) and identifies that the latency SLO is violated due to 4 hot keys. As a result, the policy engine triggers the selective replication of the 4 keys. Figure~\ref{fig-load-balance} shows the KVSs' behavior during the experiment.



Initially, all the KVSs meet the latency SLO and balance the load across \cns.
At 20 seconds, the workload switches to the highly skewed pattern, resulting in
latency SLO violations and an increase in load imbalance between \cns.
\sysname gradually increases the replication factor of the 4 keys between 30 and 90 seconds. During this period, \sysname experiences brief tail latency spikes due to the additional delay for clients to retrieve the up-to-date ownership mapping of replicated keys from the \rn, but throughput gradually increases. At 90 seconds, \sysname fully replicates the hot keys across all available \cns, 
and the throughput stabilizes. The latency SLOs are also met. \sysname was the 
only system to satisfy the SLOs; both Clover and \sysname-N constantly violate the SLOs for the highly-skewed workload.

Clover initially outperforms \sysname without selective replication and \sysname-N by almost 4\myx on the highly-skewed workload.
However, once we enable selective replication in \sysname, hot keys start becoming shared by multiple \cns at about 30-40 seconds; once all the hot keys are completely replicated, \sysname's performance stabilizes in about 1 minute and it outperforms Clover by almost 1.6\myx and \sysname-N up to 5.6\myx.
Selectively replicating hot keys in \sysname allows multiple \cns to access \dpm for the hot keys, increasing the overall throughput.
Our use of indirect pointers in accessing hot keys restricts \cns from caching values.
Hence, \sysname selectively replicates only the hottest keys while restricting \cns to cache only their shortcuts; \cns maintain exclusive ownership over non-hot keys and continue to cache their values adaptively.

Overall, our experiments highlight the benefits of selective replication with \dip for load balancing across \cns and for handling hot spots as a better alternative to shared-everything.

\begin{figure}
  \begin{center}
    \includegraphics[width=0.47\textwidth]{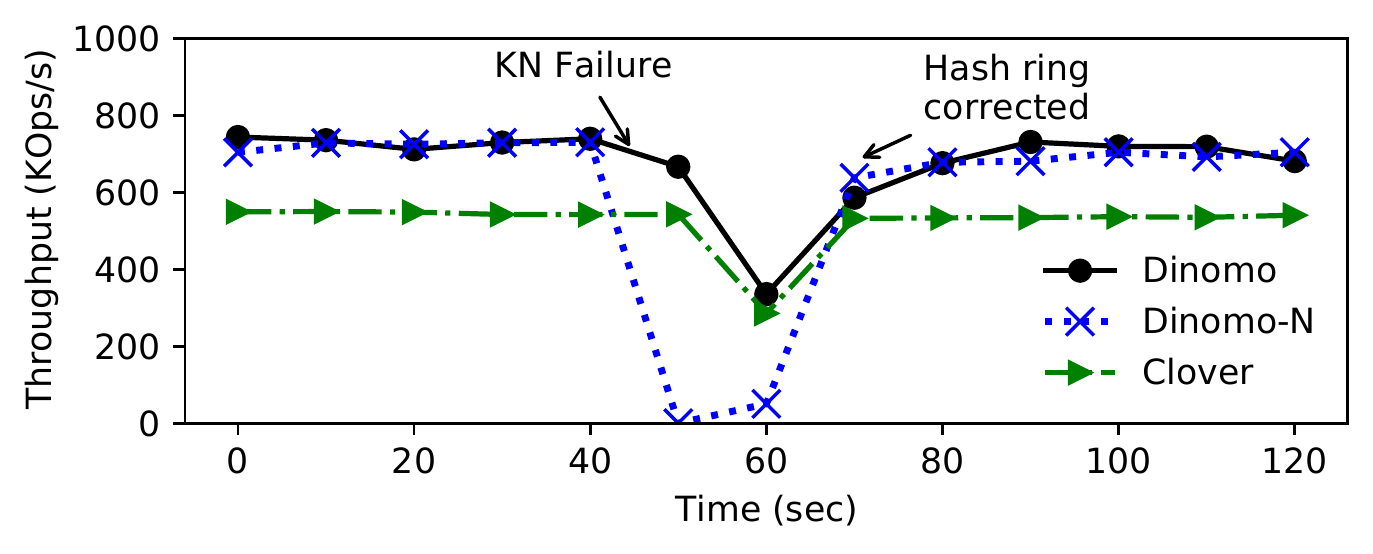}
  \end{center}
\if\removespace 1
  \vspace{-15pt}
\fi
  \caption{Throughput of \sysname, \sysname-N, and Clover over time while handling a \cn failure}{}
  \label{fig-failover-throughput}
  \if\removespace 1
  \vspace{-10pt}
  \fi
\end{figure}

\vheading{Fault Tolerance.}
Finally, we induce a KN failure to compare the resilience and elasticity of \sysname, \sysname-N and Clover.
In a cluster with 16 \cns, we run a moderate skew (Zipf 0.99) workload for 2 minutes using 8 client nodes, and simulate a \cn failure at around 40 seconds.
We simulate the failure by eliminating a randomly selected \cn. User requests are set to time out after 500ms.
We observe that \sysname quickly recovers from the \cn failure (Figure~\ref{fig-failover-throughput}).
We notice that the throughput briefly drops by 45\%, average latency increases by 1.2\myx (0.8 ms), and the tail latency increases by 1.5\myx (1.4 ms).
Upon detecting the failure, \sysname merges the pending log segments from the failed \cn and redistributes ownership across other alive \cns.
These steps take less than 109 ms.

\sysname-N, on the other hand, experiences a 20-second dip in performance at 50 seconds, where the throughput drops to 0 as it stops serving requests while reshuffling data.
The time to reorganize data takes more than 11 seconds in \sysname-N.
Clover tolerates the \cn failure elastically, showing a brief 55\% drop in its throughput.
Clover only needs to update the cluster membership of alive \cns in \rns after failures (without any data reorganization) to allow clients to retrieve the new membership after timeouts.
The time to update \rns takes less than 68 ms.

Overall, compared to \sysname-N, \sysname recovers from \cn failure faster since it is not required to reorganize data owing to the data sharing in \dip.
Similar to Clover, \sysname stabilizes its performance quickly, and satisfies all SLOs.

\section{Related work}
\label{sec:related-work}

We place our contributions in the context of relevant prior work.

\vheading{\dpm architectures.}
\dip follows the idea that just because you \emph{can} share,
    it does not mean you \emph{should} share.
This observation has been made before in other contexts.
Storage Area Networks provide storage disaggregation in a data center~\cite{khattar1999introduction},
    where volumes could be shared among hosts, but often they are not~\cite{caulfieldISCA13}.
Key-value stores provide storage disaggregation in the cloud,
    where data can be shared among nodes, but applications may
    choose not to~\cite{snowflake}.
Fine-grained logical partitioning has been proposed
    to support live reconfigurations in
    in-memory key-value stores~\cite{kulkarni17rocksteady,atulOSDI16},
    in-memory databases~\cite{elmore15squall}, and
    graph processing~\cite{xie19pragh}.
Even multiprocessor shared-memory systems sometimes forgo sharing
    of data structures among threads, choosing instead to partition data~\cite{baumann09barrelfish,delegation,lim2014mica}.
Our work demonstrates that partitioning logical ownership while
    sharing physical data and metadata in \dpm provides
    high performance and lightweight reconfigurability.

\vheading{\cp.}
Adaptive caching policies have been explored in other contexts,
    illustrating how a single cache can be used for multiple purposes
    or how a replacement policy can consider multiple behaviors.
For example, the Sprite operating system shared its memory between
    the file system buffer cache and the virtual memory
    system~\cite{nelson88spritecache}.
The Adaptive Replacement Cache (ARC) uses a replacement policy that
    balances between recency and frequency of accesses~\cite{megiddo03arc}.
In contrast to these systems, which use fixed-size cache entries
    with uniform miss penalties, \cp manages a cache where different
    types of entries (e.g., values vs. shortcuts) have different sizes
    and varying miss penalties.
The novelty of our scheme arises from a new setting (\dpm) where
    adaptivity is essential.



\section{Conclusion}
\dpm is a promising new technology for building KVSs.
Prior \dpm KVSs have had to sacrifice at least one of three desirable characteristics:
   high common-case performance, scalability, and lightweight reconfiguration. 
We present the \sysname KVS, which uses a novel combination of techniques to achieve
   these properties simultaneously, which we demonstrate through empirical evaluation.

\if\camerareadyversion 1
\begin{acks}
We would like to thank the anonymous reviewers and members of Systems and Storage Lab for their insightful comments and constructive feedback.
This work was supported by NSF CAREER \#1751277 and donations from VMware, Google, and Meta. Sekwon Lee was also supported by the Microsoft Research PhD Fellowship.
\end{acks}
\fi

\bibliographystyle{ACM-Reference-Format}
\bibliography{main}

\end{document}